\pdfoutput=1
\documentclass[12pt]{article}
\usepackage[utf8]{inputenc}
\usepackage{amsmath}
\usepackage{booktabs}
\usepackage{cite}
\usepackage{amssymb}
\usepackage[title]{appendix}
\usepackage{mathrsfs}
\usepackage{float}
\usepackage{multirow}
\usepackage{graphicx,caption,subcaption}
\usepackage[space]{grffile}
\usepackage{fullpage}
\usepackage{color}
\usepackage{soul}

\usepackage{datetime}
\newdateformat{monthyeardate}{%
  \monthname[\THEMONTH], \THEYEAR}
\date{\monthyeardate\today}

\usepackage{hyperref}

\newcommand{\overbar}[1]{\mkern1.5mu\overline{\mkern-1.5mu#1\mkern-1.5mu}\mkern 1.5mu}

\begin{document}

\renewcommand{\arraystretch}{1.3}
\thispagestyle{empty}

{\hbox to\hsize{\vbox{\noindent January 2021}}}

\noindent
\vskip2.0cm
\begin{center}

{\Large\bf $\alpha$-attractors from supersymmetry breaking}

\vglue.3in

Yermek Aldabergenov,~${}^{a,b,}$\footnote{yermek.a@chula.ac.th} Auttakit Chatrabhuti,~${}^{a,}$\footnote{auttakit.c@chula.ac.th} and Hiroshi Isono~${}^{a,}$\footnote{hiroshi.isono81@gmail.com}
\vglue.1in

${}^a$~{\it Department of Physics, Faculty of Science, Chulalongkorn University\\
Thanon Phayathai, Pathumwan, Bangkok 10330, Thailand}\\
${}^b$~{\it Institute of Experimental and Theoretical Physics, Al-Farabi Kazakh National University,
71 Al-Farabi Avenue, Almaty 050040, Kazakhstan}
\vglue.1in

\end{center}

\vglue.3in

\begin{center}
{\Large\bf Abstract}
\vglue.2in
\end{center}

We construct new models of inflation and spontaneous supersymmetry breaking in de Sitter vacuum, with a single chiral superfield, where inflaton is the superpartner of the goldstino. Our approach is based on hyperbolic K\"ahler geometry, and a gauged (non-axionic) $U(1)_R$ symmetry rotating the chiral scalar field by a phase. The $U(1)_R$ gauge field combines with the angular component of the chiral scalar to form a massive vector, and single-field inflation is driven by the radial part of the scalar. We find that in a certain parameter range they can be approximated by simplest Starobinsky-like (E-model) $\alpha$-attractors, thus predicting $n_s$ and $r$ within $1\sigma$ CMB constraints. Supersymmetry (and $R$-symmetry) is broken at a high scale with the gravitino mass $m_{3/2}\gtrsim 10^{14}$ GeV, and the fermionic sector also includes a heavy spin-$1/2$ field. In all the considered cases the inflaton is the lightest field of the model.

\newpage

\tableofcontents

\section{Introduction}

Embedding inflationary cosmology in supergravity is known to be highly non-trivial due to the restricted form of (locally) supersymmetric actions. Despite this fact, there are many successful examples of supergravity based inflation.~\footnote{See for example the reviews \cite{Yamaguchi:2011kg,Ellis:2020lnc}.} CMB observations \cite{Hinshaw:2012aka,Akrami:2018odb} so far favor single-field inflation with a concave potential and sufficiently flat inflationary plateau. Some of the simplest models with these features have scalar potentials of the form $V\sim (1-e^{a\varphi})^2$ (assuming canonical kinetic term for $\varphi$), where $\varphi$ is the inflaton field, and $a$ is some real constant. A particular choice $a=\sqrt{2/3}$ corresponds to the well-known Starobinsky model (also known as $R^2$ inflation) \cite{Starobinsky:1980te}. Within supergravity framework there exist many possible realizations of the Starobinsky model and its generalizations with arbitrary values of $a$, see e.g. Refs. \cite{Cecotti:1987sa,Ketov:2012yz,Farakos:2013cqa,Ferrara:2013rsa,Ketov:2013dfa,Antoniadis:2014oya,DallAgata:2014qsj,Ketov:2014qoa,Hasegawa:2015era,Ellis:2015kqa,Ellis:2015xna,Ellis:2017xwz,Ellis:2018xdr,Ellis:2018zya,Aldabergenov:2019aag,Aldabergenov:2020atd}. These models are known as E-model $\alpha$-attractors \cite{Kallosh:2013yoa,Kallosh:2013lkr,Cecotti:2014ipa,Linde:2015uga,Carrasco:2015uma,Ellis:2019bmm}. These can be further generalized but we will consider only the simplest E-models.

In most of the inflationary supergravity models supersymmetry (SUSY) is restored after inflation, and is often assumed to be broken by a separate field(s) in the hidden sector (e.g. Polonyi field) \cite{Nilles:1983ge,Kribs:2010md,Ferrara:2016vzg,Aldabergenov:2016dcu,Pallis:2018xmt}, which then mediates SUSY breaking to a supersymmetric Standard Model via Planck-suppressed interactions. However there is an alternative, more economical possibility -- inflaton itself may be responsible for SUSY breaking (sgoldstino inflation) if its $F$-term is non-vanishing at the minimum of the potential \cite{Izawa:1997jc,AlvarezGaume:2010rt,AlvarezGaume:2011xv,Achucarro:2012hg,Dalianis:2014aya,Schmitz:2016kyr,Antoniadis:2016aal,Antoniadis:2017gjr}.

In this work we further study sgoldstino inflation scenarios, by focusing on the models with gauged $R$-symmetry. In particular, we take a class of SUSY breaking models introduced in Ref. \cite{Aldabergenov:2019ngl}, where the (complex) scalar field parametrizes an $SU(1,1)/U(1)$ K\"ahler space, and both supersymmetry and $R$-symmetry can be spontaneously broken in de Sitter vacuum. The models are given by the following K\"ahler potential and superpotential,~\footnote{In Ref. \cite{Antoniadis:2017gjr} a similar type of inflationary models was considered, with gauged $U(1)_R$ phase symmetry which is broken (together with SUSY) at the minimum. The main difference is that the model of \cite{Antoniadis:2017gjr} dealt with almost canonical K\"ahler potential $K=Z\overbar{Z}+...$ which lead to small-field hilltop inflation. Here our models will be shown to be large-field after canonical parametrization.}
\begin{equation}
    K=-\alpha\log(1-Z\overbar{Z})~,~~~W=\mu Z^n~,\label{KW}
\end{equation}
where $Z$ is the chiral scalar field playing the role of sgoldstino. The $U(1)_R$ phase symmetry $Z\rightarrow e^{-iq_z\theta}Z$, where $q_Z$ is the corresponding charge and $\theta$ is the transformation parameter, is promoted to a gauge symmetry, which generates a $D$-term contribution to the scalar potential. The number $n$ in \eqref{KW} is given by $n=q/q_Z$ where $q$ is the $U(1)_R$ charge of the superpotential. Various choices of $\alpha$ and $n$ allow for the spontaneous SUSY and R-symmetry breaking \cite{Aldabergenov:2019ngl}, with the integer values of $\alpha$ in the range $1\sim 7$ motivated by string theory compactifications \cite{Duff:2010ss,Duff:2010vy,Ferrara:2016fwe}. Some of the models were shown in \cite{Aldabergenov:2019ngl} to have a double-well potential (e.g. when $\alpha=1$ and $\alpha=4$), so that in principle they can support hilltop inflation. But in the simplest model \eqref{KW} the potential is not flat enough around the maximum, so that the spectral index is incompatible with observations. On the other hand $\alpha=2$ and $\alpha=3$ allows for a flat scalar potential with a tunable cosmological constant (no-scale de Sitter model). This is also not suitable for slow-roll inflation. 

Our goal is to investigate the possibility of viable inflation in the model \eqref{KW} (while keeping supersymmetry breaking Minkowski/de Sitter vacuum) by introducing simple modifications to the model allowed by the $R$-symmetry. More specifically, we add a correction term $f(Z\overbar{Z})$ to the K\"ahler potential,
\begin{equation}
    K=-\alpha\log(1-Z\overbar{Z}-f(Z\overbar{Z}))~,
\end{equation}
and show that when $\alpha=2$ and $\alpha=3$ it suffices to consider the leading-order correction $f=\beta (Z\overbar{Z})^2$ in order to realize slow-roll (hilltop) inflation, while in the superpotential \eqref{KW} we can set $n=1$.

This paper is organized as follows. In Section \ref{sec_the_model} we derive our master Lagrangian, discuss the corresponding hyperbolic K\"ahler geometry, and make analytical estimations for the inflationary parameters in slow-roll regime. In Sections \ref{sec_alpha_2} and \ref{sec_alpha_3} we study in detail two viable models defined by the choice of $\alpha$, one with $\alpha=2$ and the other with $\alpha=3$. In each case we derive the inflationary observables, bosonic mass spectrum and SUSY breaking parameters, and make a comparison with simple $\alpha$-attractor models. In Section \ref{sec_fermion} we discuss the physical fermionic spectrum in which the goldstino vanishes, leaving a single heavy spin-$1/2$ field. We comment on the restrictions on the parameter space that can be put by the Swampland Distance Conjecture in Section \ref{sec_SDC}. Section \ref{sec_concl} is devoted for conclusion and discussion.

Throughout the paper we use the observed values of the spectral index, tensor-to-scalar ratio, and the amplitude of scalar perturbations provided by PLANCK collaboration \cite{Akrami:2018odb},
\begin{gather}
    n_s=0.9649\pm 0.0042~{\rm (1\sigma)}~,~~~r<0.064~{\rm (2\sigma)}~,\nonumber\\
    \log(10^{10}A_s)=2.975\pm 0.056~{\rm (1\sigma)}~\Rightarrow~A_s\approx 1.96\times 10^{-9}~\label{nsrA_obs}.
\end{gather}

\section{The model}\label{sec_the_model}

As outlined in Introduction, we study the models given by
\begin{equation}
K=-\alpha\log(1-|Z|^2-\beta|Z|^4)~,~~~W=\mu Z~,\label{Kahler_beta}
\end{equation}
and a trivial gauge kinetic function $h=1$.~\footnote{In Ref.\,\cite{Antoniadis:2017gjr} it was shown that a field-dependent gauge-kinetic function $h(Z)$ can be used to cancel anomalies due to the $R$-charged fermions (via Green-Schwarz mechanism). However, the field-dependence of $h(Z)$ was shown to be negligible so that it can be approximated as $h(Z)\approx 1$. This result can also be applied to our models, since we have the same field content and $R$-charge assignment.} We consider gauged $U(1)_R$ symmetry with the gauge coupling $g$, that rotates the chiral field $Z$ by a phase.

In the bosonic sector, this leads to the Lagrangian (see Appendix for the derivation)
\begin{equation}
    e^{-1}{\cal L}=\tfrac{1}{2}R-K_{Z\overbar{Z}}D_mZ\overbar{D^mZ}-\tfrac{1}{4}F_{mn}F^{mn}-V_F-V_D~,\label{L_beta}
\end{equation}
where the K\"ahler metric and the covariant derivative are given by 
\begin{gather}
    K_{Z\overbar{Z}}=\alpha\frac{1+4\beta|Z|^2-\beta|Z|^4}{(1-|Z|^2-\beta|Z|^4)^2}~,~~~D_mZ=\partial_mZ+igA_mZ~,\label{Kahler_metric_beta}
\end{gather}
while the scalar potential is the sum $V=V_F+V_D$, with the $F$- and $D$-term contributions given by
\begin{gather}
    V_F=\frac{\mu^2}{(1-|Z|^2-\beta|Z|^4)^\alpha}\left\{\frac{\left(1+(\alpha-1)|Z|^2+\beta(2\alpha-1)|Z|^4\right)^2}{\alpha(1+4\beta|Z|^2-\beta|Z|^4)}-3|Z|^2\right\}~,\label{VF_beta}\\
    V_D=\frac{g^2}{2}\left\{\frac{1+(\alpha-1)|Z|^2+\beta(2\alpha-1)|Z|^4}{1-|Z|^2-\beta|Z|^4}\right\}^2~,\label{VD_beta}
\end{gather}
where $Z=0$ is the symmetric point of the potential which we assume to be the local maximum, and its absolute value is bounded by $1-|Z|^2-\beta |Z|^4>0$ due to the singularity in the K\"ahler metric \eqref{Kahler_metric_beta}. The tensor $F_{mn}$ is the field strength of the $U(1)_R$ gauge field $A_m$. 

It is convenient to use the parametrization $Z=z\,e^{-i\zeta}$, where $z$ is the absolute value of $Z$, while $\zeta$ is its angular (Goldstone) mode. If the vacuum expectation value (VEV) of $Z$ is non-zero, $A_m$ will acquire physical mass proportional to $\langle Z\rangle^2$ while the Goldstone mode can be gauged away. Given by Eqs.\,\eqref{L_beta}--\eqref{VD_beta} is our master Lagrangian.

The parameters $\alpha$ and $\beta$ control the geometry of the target space, with the K\"ahler curvature
\begin{equation}
    R_K=-\frac{2}{\alpha}(1+2\beta)+\frac{12\beta}{\alpha}(1+4\beta)z^2+{\cal O}(z^4)~,
\end{equation}
which reduces to the Poincar\'e disk value, $-2/\alpha$, in the limit $\beta=0$ (in this limit all the higher-order terms vanish). In fact, in the following sections we will show that in order to realize inflation we need $|\beta|\ll 1$, in which case $z$ is bounded from above as $z\lessapprox 1$ and for any $z$ between zero and one we have $R_K\approx -2/\alpha$.

For each choice of the geometry (set by $\alpha$ and $\beta$), the parameters $\mu$ and $g$ control the corresponding scalar potential and vacuum structure. Without loss of generality we can assume that all the parameters are real. In addition, we consider only integer values of $\alpha$ of order one, motivated by string constructions. We find that when $\alpha=2$ and $\alpha=3$, the K\"ahler potential \eqref{Kahler_beta} is suitable for hilltop inflation (near $z=0$) and stable de Sitter or Minkowski minimum (away from $z=0$). Once $\alpha$ is set, our goal is to identify the appropriate parameter region of $\beta$, $\mu$ and $g$. To do this we use slow-roll analysis of the potential near the symmetric point $z=0$, together with the vacuum equations
\begin{equation}
    V'=0~,~~~V=V_0~,\label{vacuum_eqs}
\end{equation}
at the minimum, where $V_0$ is the cosmological constant (of order $10^{-120}$) and the prime denotes the derivative with respect to the canonically parametrized scalar. Such a small cosmological constant requires very precise cancellation between positive contribution from $F$- and $D$-terms, and negative contribution from the gravitino mass (see Eq. \mbox{\eqref{V_FD_app}}).

The canonical scalar, which we denote as $\varphi$, must satisfy
\begin{equation}
    e^{-1}{\cal L}\supset -G(z)\partial_m z\partial^m z=-\tfrac{1}{2}\partial_m\varphi\partial^m\varphi~,
\end{equation}
where we introduced the notation
\begin{equation}
    G(z)\equiv K_{Z\overbar{Z}}(z)=\alpha\frac{1+4\beta z^2-\beta z^4}{(1-z^2-\beta z^4)^2}~,\label{G_def}
\end{equation}
using the K\"ahler metric \eqref{Kahler_metric_beta}. Then $\varphi$ as a function of $z$ can be found by solving
\begin{equation}
    \frac{d\varphi}{dz}=\sqrt{2G(z)}~,\label{varphi_z_alpha2}
\end{equation}
and $z(\varphi)$ is found by inverting the solution.

For the slow-roll analysis we define the ``potential" slow-roll parameters as
\begin{align}
    \epsilon_V&\equiv\frac{1}{2}\left(\frac{V_\varphi}{V}\right)^2=\frac{1}{2}\left(\frac{V_z z_\varphi}{V}\right)^2~,\\
    \eta_V&\equiv\frac{V_{\varphi\varphi}}{V}=\frac{V_{zz}z_\varphi^2+V_z z_{\varphi\varphi}}{V}~,
\end{align}
where the indices $\varphi$ and $z$ denote the respective derivatives. The derivative $z_\varphi$ (as function of $z$) is found from Eq.\,\eqref{varphi_z_alpha2}. 

Inflation starts near $z=0$ (local maximum), and $z$ slowly rolls down to a stable near-Minkowski minimum. Using the scalar potential \eqref{VF_beta}$+$\eqref{VD_beta}, the slow-roll parameters can be expanded near $z=0$ as
\begin{align}
    \epsilon_V&=\frac{4}{\alpha}\left(\frac{\alpha^2g^2-2\mu^2-4\beta\mu^2}{\alpha g^2+2\mu^2}\right)^2 z^2+{\cal O}(z^4)~,\label{epsilon_V_beta}\\
    \eta_V&=\frac{2(\alpha^2g^2-2\mu^2-4\beta\mu^2)}{\alpha(\alpha g^2+2\mu^2)}+{\cal O}(z^2)~,\label{eta_V_beta}
\end{align}
and the corresponding spectral index reads
\begin{equation}
    n_s\simeq 1+2\eta_V-6\epsilon_V=\frac{5\alpha^2g^2+2(\alpha-4-8\beta)\mu^2}{\alpha(\alpha g^2+2\mu^2)}+{\cal O}(z^2)~,\label{ns_V_beta}
\end{equation}
where at the leading order in $z$, the contribution of $\epsilon_V$ can be ignored, so that the maximized value of the spectral index, $n_s^{\rm max}\equiv n_s|_{z=0}$ can be obtained. We can then impose the following condition,
\begin{equation}
    0.9649\leq n_s^{\rm max}<1~,\label{ns_range_beta}
\end{equation}
where we used the $1\sigma$ value \eqref{nsrA_obs} as the lower limit (for simplicity we ignore the uncertainty in the values). Because the value of $n_s(z)$ drops as the inflaton rolls down the hill, if $n_s^{\rm max}<0.9649$ it can never reach this observed value. But if $n_s^{\rm max}>0.9649$, $n_s(z)$ will inevitably cross the value $0.9649$ as $z$ departs from the origin and moves towards the minimum (although this does not strictly guarantee that at the horizon exit $n_s=0.9649$). At the same time $n_s^{\rm max}$ cannot be larger than one because we require that the potential is concave near the origin ($\eta_V$ must be negative).

Let us estimate the limits of small $g$ and small $\mu$. If $g^2\ll\mu^2$ then from \eqref{ns_V_beta} we have $n_s^{\rm max}\simeq 1-4(1+2\beta)/\alpha$, so that imposing the condition \eqref{ns_range_beta} leads to $\alpha/(1+2\beta)\gtrsim{\cal O}(100)$ which is incompatible with our assumption $\alpha\sim {\cal O}(1)$ unless $\beta\sim -1/2$. However, this value of $|\beta|$ is too large to support the desired shape of the potential, as we will show briefly. On the other hand, if $g^2\gg\mu^2$ then $n_s^{\rm max}\simeq 5$ which is incompatible with observations. This leads to the conclusion that the values of $g$ and $\mu$ must be of the similar order of magnitude in order to describe inflation and a stable Minkowski/de Sitter minimum.

As for the inflationary Hubble scale, it can be estimated as
\begin{equation}
    H\simeq\sqrt{\frac{V|_{z\rightarrow 0}}{3}}~,~~~{\rm with}~~V|_{z\rightarrow 0}\simeq \frac{\mu^2}{\alpha}+\frac{g^2}{2}~.\label{alpha2_Hubble_scale}
\end{equation}
Of course, the observable inflation (e.g. the last $\sim 60$ e-folds) may start away from $z=0$, but the above estimate is still correct because the potential must be flat near the origin, i.e. its height cannot change significantly (compared to the $z=0$ value) at the point where the observable inflation starts.

In what follows let us study $\alpha=2$ and $\alpha=3$ in more detail and provide specific working examples for each case.

\section{\texorpdfstring{$\alpha=2$}{} case}\label{sec_alpha_2}
\subsection{Identifying the relevant parameter region}

As shown in \cite{Aldabergenov:2019ngl}, when $\alpha=2$ (and $\beta=0$) we can obtain no-scale de Sitter supergravity after imposing the relation $\mu^2=2g^2$ where the resulting positive cosmological constant leads to exponential expansion of the universe. If we want to apply this to slow-roll inflation we need to break the no-scale de Sitter structure, in order to introduce a stable near-Minkowski minimum. Therefore it is useful to introduce the parametrization
\begin{equation}
    \mu^2\equiv 2g^2(1+\omega)~,\label{alpha2_omega_def}
\end{equation}
where the new parameter $\omega$ (as well as $\beta$) measures the deviation from the no-scale de Sitter case and -- as will be shown -- this deviation must be small ($|\omega|\ll 1,|\beta|\ll 1$) for slow-roll inflation. It can be seen that after substituting the above relation in Eq.\,\eqref{VF_beta}, the parameter $g^2$ becomes an overall factor in the total scalar potential $V=V_F+V_D$, and thus does not affect the shape of the potential, while it sets the amplitude of inflationary scalar perturbations as well as the Hubble scale. This can be seen from Eq.\,\eqref{alpha2_Hubble_scale} (substituting \eqref{alpha2_omega_def}),
\begin{equation}
    3H^2\simeq V|_{z\rightarrow 0}=\frac{g^2}{2}(3+2\omega)~.
\end{equation}

The spectral index \eqref{ns_V_beta} in terms of $\omega$ reads
\begin{equation}
    n_s\simeq 1-4\frac{\omega+2(1+\omega)\beta}{3+2\omega}+{\cal O}(z^2)~,
\end{equation}
while the condition \eqref{ns_range_beta} for $n_s^{\rm max}$ becomes
\begin{equation}
    0<\frac{\omega+2(1+\omega)\beta}{3+2\omega}\leq 0.0088~.
\end{equation}
It can be seen that this condition can be satisfied even if one of the two parameters ($\beta,\omega$) is zero. However both parameters are needed to support a stable Minkowski or de Sitter minimum.

\begin{figure}
\centering
\begin{subfigure}{.49\textwidth}
  \centering
  \includegraphics[width=.9\linewidth]{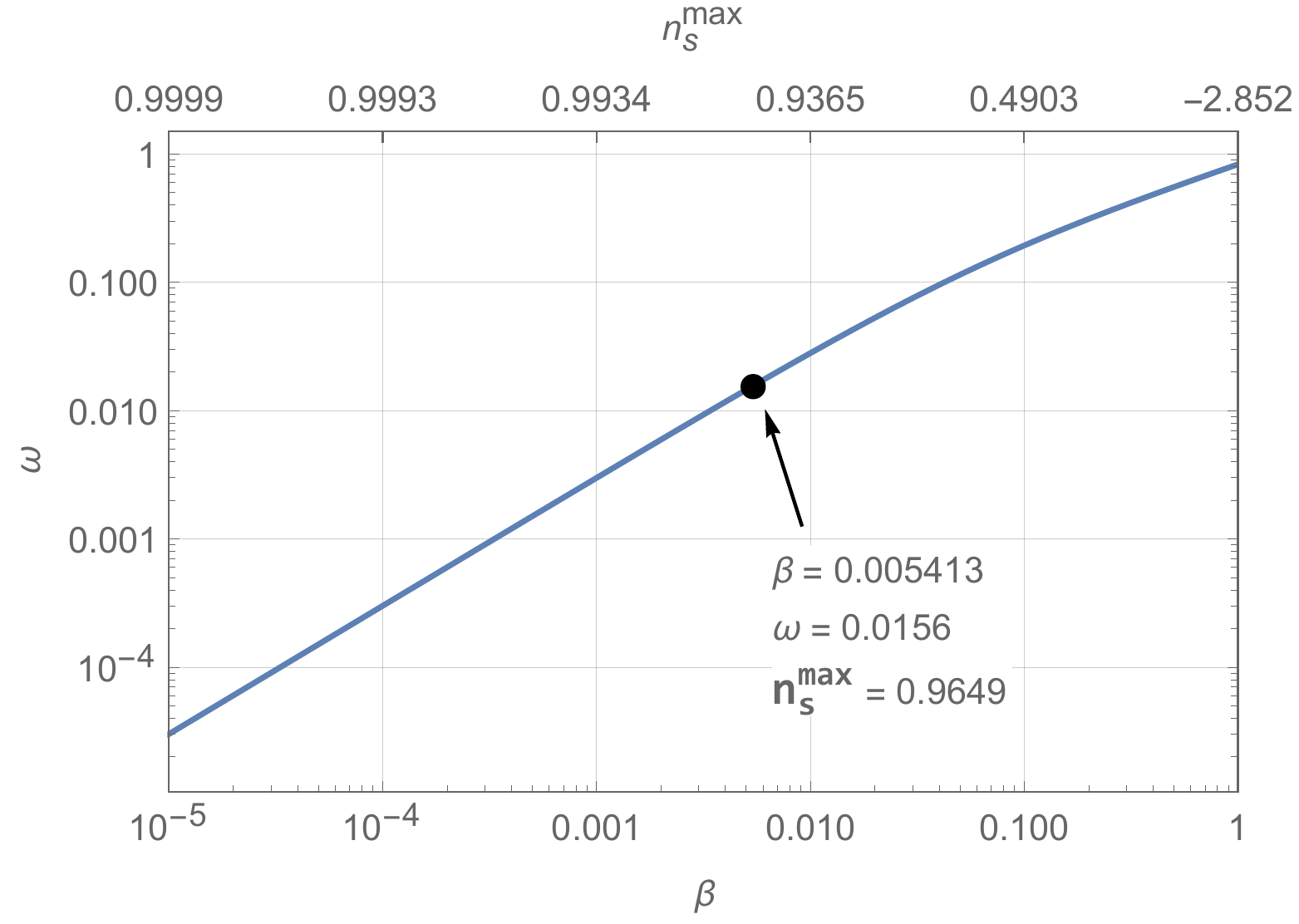}
\end{subfigure}
\begin{subfigure}{.49\textwidth}
  \centering
  \includegraphics[width=.88\linewidth]{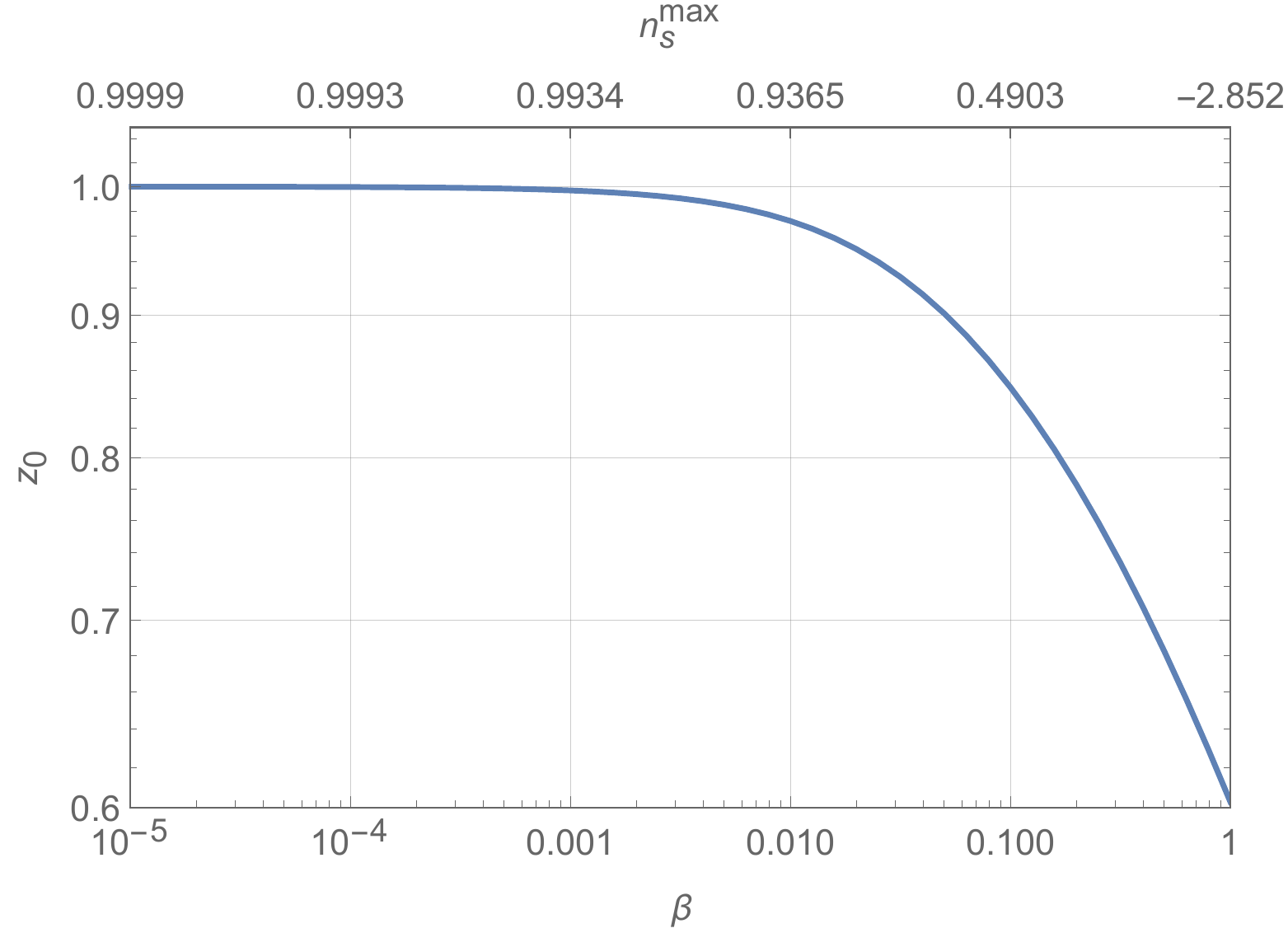}
\end{subfigure}
\captionsetup{width=.9\linewidth}
\caption{Solution to the vacuum equations $V_z=V=0$ ($\alpha=2$ case) as functions $\omega(\beta)$ (left) and $z_0(\beta)$ (right). At the top-side of the plots we provide the values of $n_s^{\rm max}$ at reference values of $\beta$.}
\label{Fig_alpha2_omega_beta}
\end{figure}

Let us now explicitly find a parameter range that allows for the stable near-Minkowski minimum by numerically solving the vacuum equations \eqref{vacuum_eqs}. Because the cosmological constant must be very small, we use the Minkowski limit, $V_0=0$. As $g^2$ is an overall factor in the potential, it is not fixed by the vacuum equations, and therefore we have three unknowns -- $\beta$, $\omega$, and $z_0\equiv\langle z\rangle$ -- and the two equations $V_z=V=0$ (assuming that $z_\varphi$ is non-zero away from the origin). We solve these by varying $\beta$ in the range $10^{-5}\leq\beta\leq 1$ and finding the corresponding values of $\omega$ and $z_0$. The results are shown in Figure \ref{Fig_alpha2_omega_beta}, together with the corresponding values of $n_s^{\rm max}$. It can be seen that the requirement $n_s^{\rm max}\gtrsim 0.9649$ leads to the upper limit on both $\beta$ and $\omega$,
\begin{equation}
    \beta\lesssim 0.0054~,~~~\omega\lesssim 0.0156~.
\end{equation}
On the other hand if we want to retain the flatness of the potential near the origin, while also having a stable Minkowski minimum away from the origin, $\beta$ and $\omega$ must be positive and non-zero. De Sitter uplifting of the minimum can easily be achieved by requiring $V=V_0>0$ at the minimum -- this will then modify the solution of Figure \ref{Fig_alpha2_omega_beta}. Of course for $V_0\sim 10^{-120}$ this modification is negligible.

\subsection{Explicit example}

In order to demonstrate inflationary solution, let us consider the following example,
\begin{equation}
    \beta=10^{-3}~,~~~\omega=2.977\times 10^{-3}~,\label{beta_omega_eg}
\end{equation}
where the parameters are within the allowed region, and taken from the vacuum solution of Figure \ref{Fig_alpha2_omega_beta}.

\begin{figure}
\centering
\begin{subfigure}{.49\textwidth}
  \centering
  \includegraphics[width=.9\linewidth]{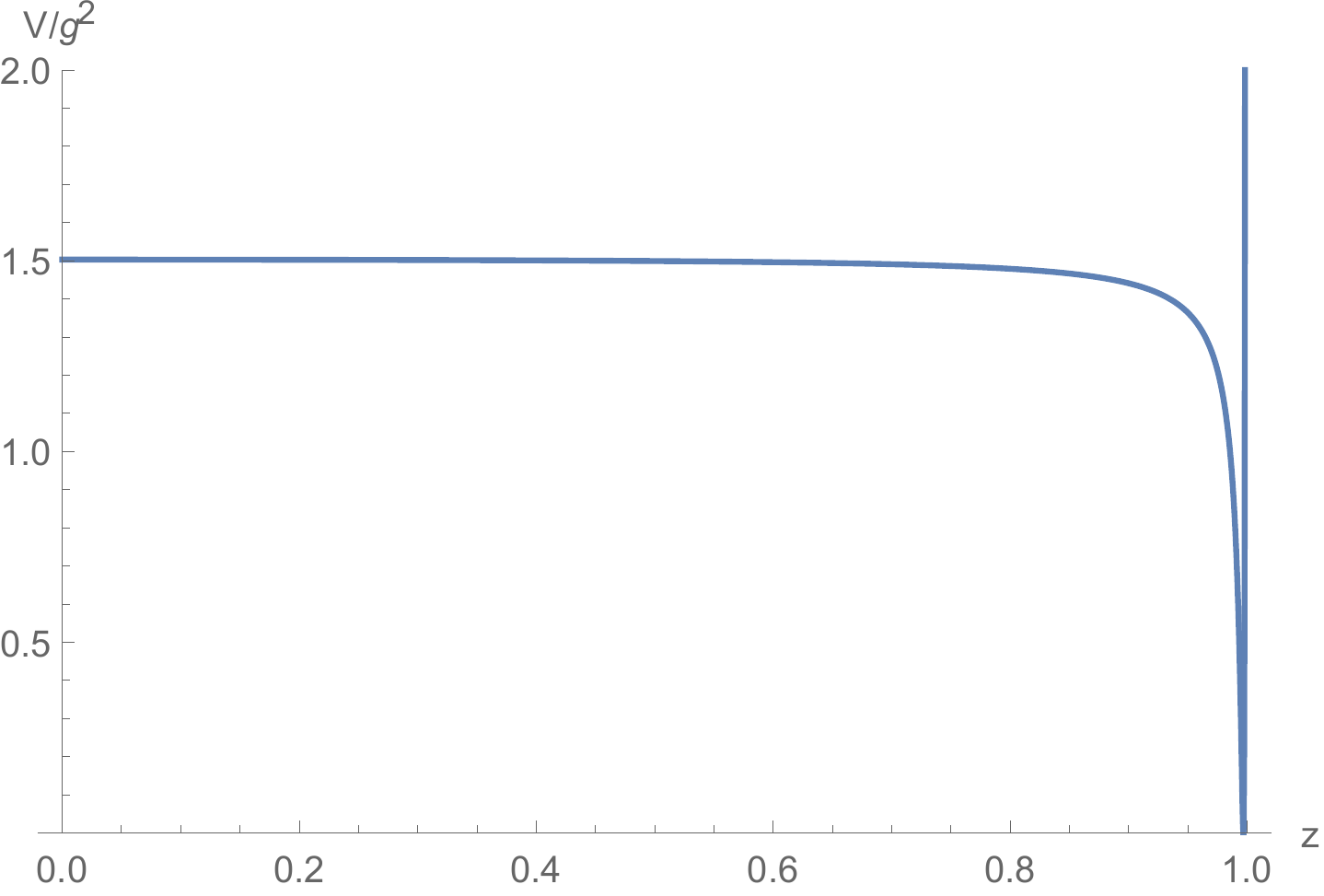}
\end{subfigure}
\begin{subfigure}{.49\textwidth}
  \centering
  \includegraphics[width=.88\linewidth]{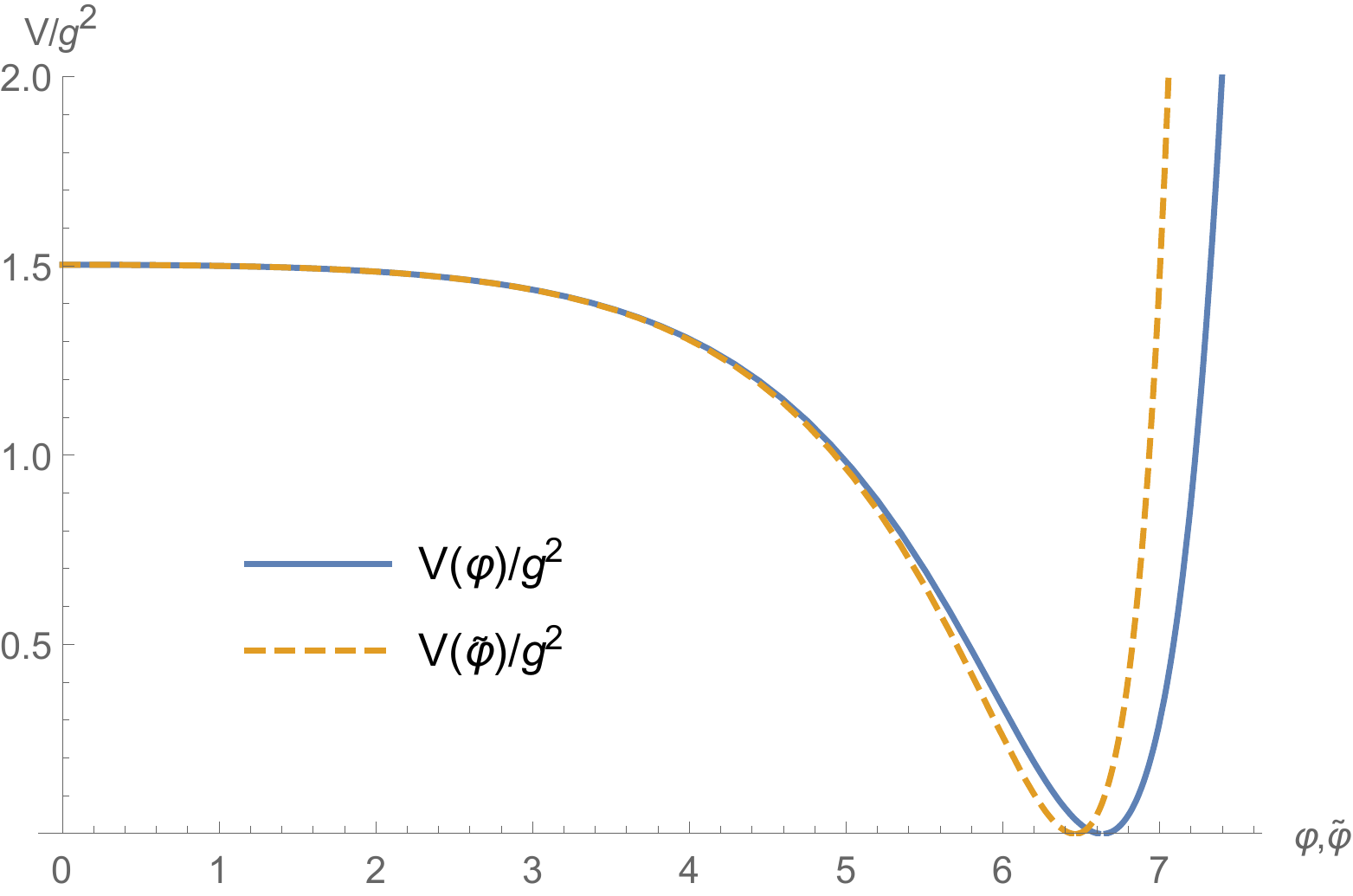}
\end{subfigure}
\captionsetup{width=.9\linewidth}
\caption{Left: the scalar potential \eqref{VF_beta}$+$\eqref{VD_beta}. Right: the same potential when using the canonical parametrization $z(\varphi)$ found by numerically inverting the function \eqref{varphi_z_alpha2} -- shown as the solid curve. The dashed curve represents the parametrization $z=\tanh{\left(\tilde{\varphi}/\sqrt{2\alpha}\right)}$ (canonical in $\beta=0$ limit). The parameters used are $\alpha=2$, $\beta=10^{-3}$, and $\omega=2.977\times 10^{-3}$.}
\label{Fig_alpha2_V}
\end{figure}

First, let us show the scalar potential \eqref{VF_beta}$+$\eqref{VD_beta} in terms of the canonical scalar $\varphi$, by numerically obtaining the function $z(\varphi)$ as the inverse of the solution to Eq.\,\eqref{varphi_z_alpha2}. After fixing $\alpha=2$, $\beta=10^{-3}$, and $\omega=2.977\times 10^{-3}$, we plot the (non-canonical) potential $V(z)$ (Figure \ref{Fig_alpha2_V}, left) and the same potential in terms of the canonical scalar $\varphi$ (Figure \ref{Fig_alpha2_V}, right) for side-by-side comparison. For reference we also plot the potential using the parametrization $z=\tanh{\left(\tilde{\varphi}/\sqrt{2\alpha}\right)}$ (orange dashed curve) which is the canonical parametrization in the pure Poincar\'e disk case with $\beta=0$, and thus corresponds to the leading-order approximation of the $\beta$-expansion of \eqref{varphi_z_alpha2}. Although any change in $z$ is sub-Planckian, the canonical scalar $\varphi$ may change by super-Planckian values during inflation (large-field inflation).

The potential $V(\varphi)$ is symmetric under the sign flip $\varphi\rightarrow -\varphi$ and has the local maximum at $\varphi=0$ (symmetric point). While $z$ has an upper limit determined by the singularity in the K\"ahler metric, the canonical scalar $\varphi$ is not bounded and can go to infinity where the potential behaves like $V|_{\varphi\rightarrow\pm\infty}\rightarrow\infty$.

Next, we estimate the inflationary observables at the horizon exit by using the potential slow-roll parameters \eqref{epsilon_V_beta}, \eqref{eta_V_beta}, as well as the number of e-folds approximated as
\begin{equation}
    \Delta N\simeq\int_{\varphi_i}^{\varphi_f}\frac{d\varphi}{\sqrt{2\epsilon_V}}=\int_{z_i}^{z_f}\frac{dz}{\sqrt{2\epsilon_V}z_\varphi}~,
\end{equation}
where $\Delta N$ is the e-folds number between $z=z_i$ and $z=z_f$. We put $\Delta N=60$, and $z_i$ as the value at the horizon exit, while $z_f$ is the value at the end of inflation defined by $\epsilon_V=1$. Setting $\beta=10^{-3}$ and $\omega=2.977\times 10^{-3}$, the observables are estimated as
\begin{equation}
    n_s\approx 0.9642~,~~~r\approx 0.0017~.
\end{equation}

Let us compare these results with the predictions of the equations of motion. The relevant part of the Lagrangian reads (see Eqs.\,\eqref{Kahler_beta}--\eqref{VD_beta})
\begin{equation}
    e^{-1}{\cal L}=\tfrac{1}{2}R-G(z)\partial_m z\partial^m z-V(z)+...~,
\end{equation}
with $G(z)$ and $V(z)$ given by Eqs.\,\eqref{G_def} and \eqref{VF_beta}\eqref{VD_beta}, and we set $\alpha=2$, $\mu^2=2g^2(1+\omega)$, $\beta=10^{-3}$ and $\omega=2.977\times 10^{-3}$.

Ignoring the dependence of $z$ on the spatial coordinates, and using Friedmann--Lema\^itre--Robertson--Walker (FLRW) metric $g_{mn}={\rm diag}(-1,a^2,a^2,a^2)$, where $a=a(t)$ is the time-dependent scale factor, we find the inflaton EoM,
\begin{equation}
    \ddot{z}+3H\dot{z}+\frac{G'}{2G}\dot{z}^2+\frac{V'}{2G}=0~,\label{KG_alpha2}
\end{equation}
where the dot stands for time derivative, the prime denotes the derivative with respect to $z$, and $H\equiv\dot{a}/a$ is the Hubble function. The Friedmann equations read
\begin{align}
    3H^2-G\dot{z}^2-V&=0~,\label{Friedmann1_alpha2}\\
    \dot{H}+G\dot{z}^2&=0~.\label{Friedmann2_alpha2}
\end{align}

For numerical solution it is convenient to use the normalized time $\tilde{t}\equiv gt$ which leads to the rescaled Hubble function $\tilde{H}=H/g$. When using $\tilde{t}$, the equations of motion become independent of $g$ (recall that $V$ contains $g^2$ as an overall factor).

\begin{figure}
\centering
  \includegraphics[width=.45\linewidth]{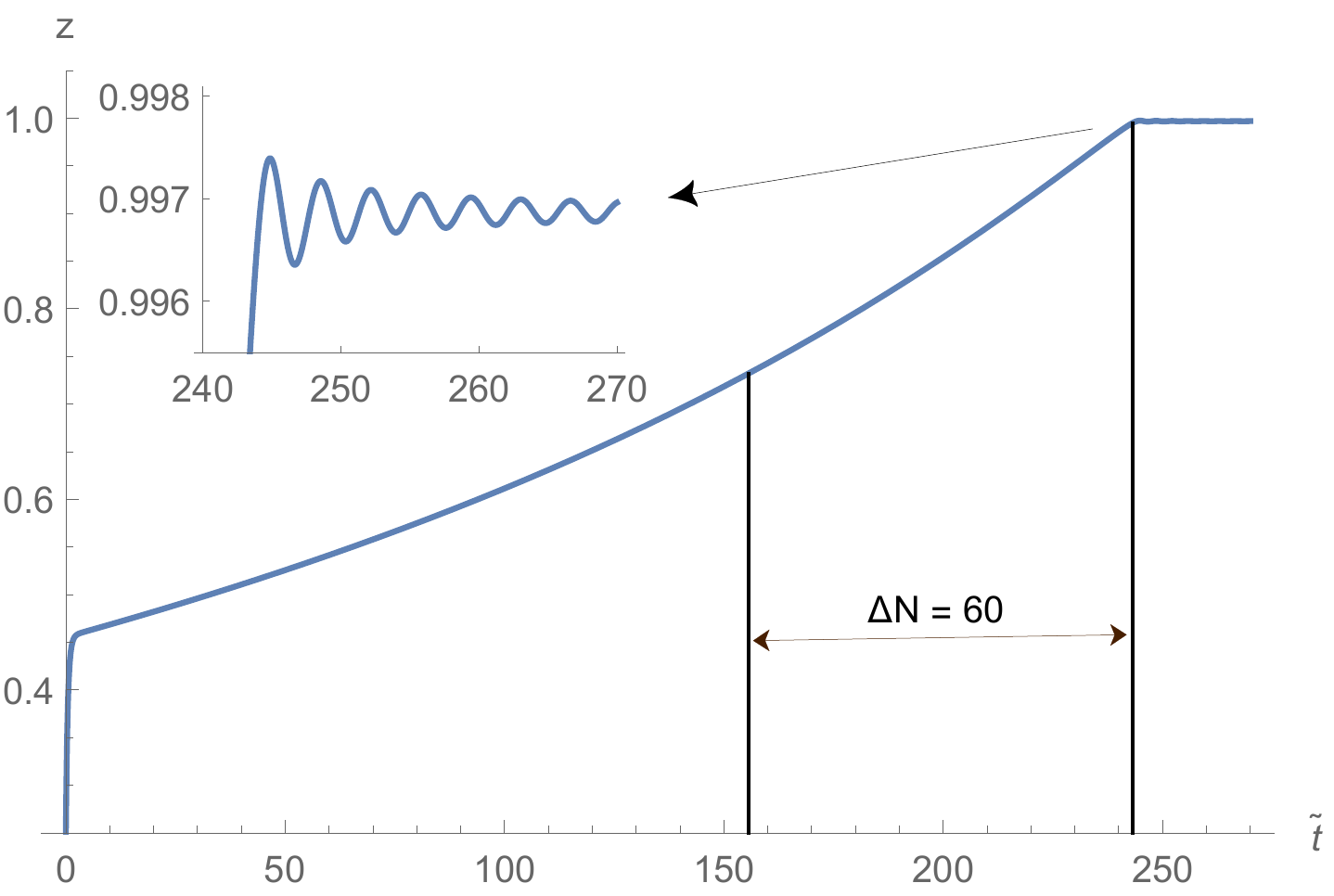}
\captionsetup{width=.9\linewidth}
\caption{The inflationary solution $z(\tilde{t})$ to the equations of motion \eqref{KG_alpha2} and \eqref{Friedmann1_alpha2} with the initial conditions $z(0)=1/4$, $\dot{z}(0)=1/2$. The vertical lines represent the last $60$ e-folds of inflation.}
\label{Fig_alpha2_zsol}
\end{figure}

\begin{figure}
\centering
\begin{subfigure}{.49\textwidth}
  \centering
  \includegraphics[width=.9\linewidth]{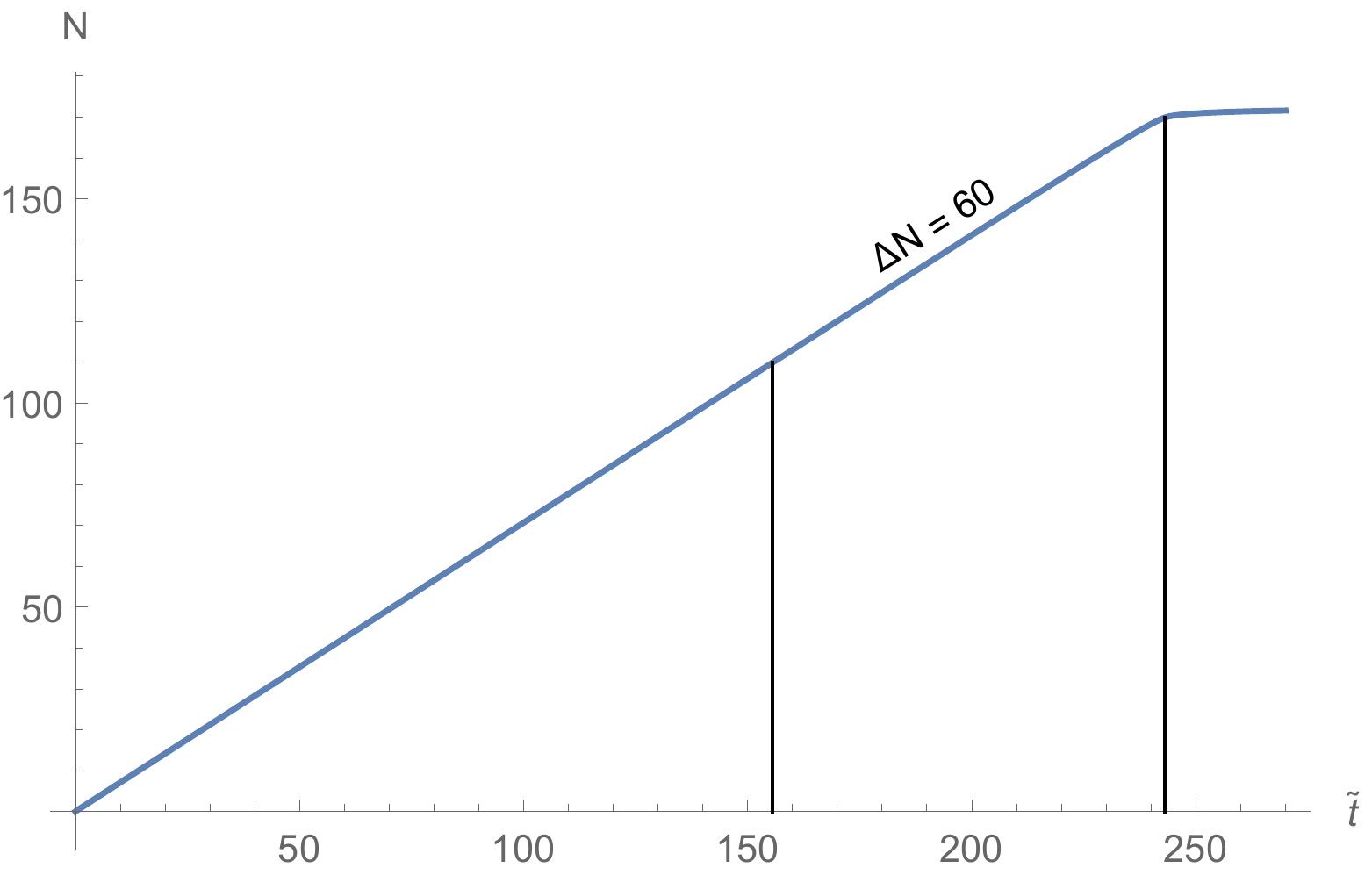}
\end{subfigure}
\begin{subfigure}{.49\textwidth}
  \centering
  \includegraphics[width=.88\linewidth]{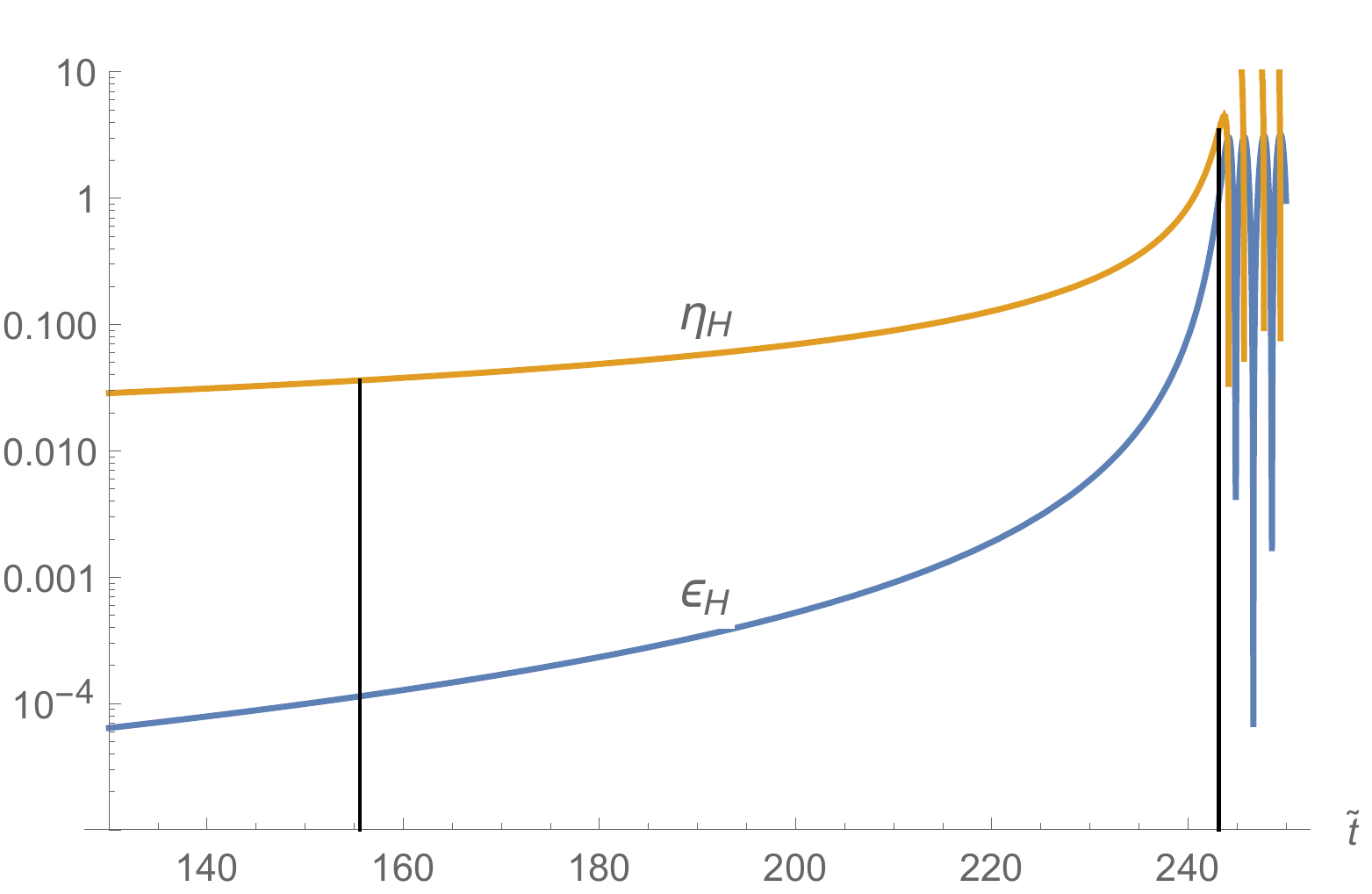}
\end{subfigure}
\captionsetup{width=.9\linewidth}
\caption{The number of e-folds (left) and the slow-roll parameters (right) for the solution to Eqs.\,\eqref{KG_alpha2} and \eqref{Friedmann1_alpha2}. The vertical lines represent the last $60$ e-folds of inflation.}
\label{Fig_alpha2_NSR}
\end{figure}

We plot the inflationary solution of Eqs.\,\eqref{KG_alpha2} and \eqref{Friedmann1_alpha2} in Figure \ref{Fig_alpha2_zsol}, where we have set the initial conditions as $z(0)=1/4$ and $\dot{z}(0)=1/2$.~\footnote{As usual, the inflationary solutions for different initial conditions join the attractor solution on $\dot{z}$--$z$ plane, so that a wide range of the initial conditions leads to the same inflationary predictions, provided the solution joins the attractor early enough to produce the required amount of inflation.} In Figure \ref{Fig_alpha2_NSR} we show the number of e-folds defined as the solution to $\dot{N}=H$ with the initial condition $N(0)=0$, as well as the Hubble slow-roll parameters defined as
\begin{equation}
    \epsilon_H\equiv -\frac{\dot{H}}{H^2}~,~~~\eta_H\equiv\frac{\dot{\epsilon}_H}{H\epsilon_H}~.
\end{equation}
On the left-side plot of Figure \ref{Fig_alpha2_NSR} the vertical lines denote the beginning and end of the last $60$ e-folds of inflation, with inflation ending when $\epsilon_H=1$.

As regards the spectral index and tensor-to-scalar ratio, when using the Hubble slow-roll parameters they can be expressed as
\begin{equation}
    n_s\simeq 1-\eta_H-2\epsilon_H~,~~~r\simeq 16\epsilon_H~.
\end{equation}
Using our solution, at the beginning of the last $60$ e-folds they are evaluated as
\begin{equation}
    n_s=0.9638~,~~~r=0.0018~,
\end{equation}
where $n_s$ is slightly smaller than the value from the slow-roll analysis using $\epsilon_V$ and $\eta_V$, and $r$ is slightly larger.

The amplitude of the scalar perturbation during slow-roll can be estimated as
\begin{equation}
    A_s\simeq \frac{H^2}{8\pi^2\epsilon_H}\simeq  \frac{V}{24\pi^2\epsilon_H}~.
\end{equation}
Finally, equating $A_s$ (at the beginning of the last $60$ e-folds) to the CMB value \eqref{nsrA_obs}, we obtain the value of the gauge coupling $g\approx 5.96\times 10^{-6}$.

\subsection{Relation to \texorpdfstring{$\alpha$}{}-attractors}

Here we will consider the scalar potential of our $\alpha=2$ models for different choices of $\beta$ (and the corresponding values of $\omega$), for the canonically normalized inflaton, and show the parameter range that asymptotically leads to the $\alpha$-attractor regime.

First, let us introduce the relevant subclass of $\alpha$-attractors. As mentioned in the introduction, we are interested in the simplest E-model $\alpha$-attractors, first constructed in Ref. \cite{Kallosh:2013yoa} using superconformal framework, and which lead to the universal prediction for the inflationary observables,
\begin{equation}
    n_s\simeq 1-\frac{2}{\Delta N}~,~~~r\simeq\frac{4\alpha}{\Delta N^2}~,
\end{equation}
where $\Delta N$ is the observable number of e-folds (we take $\Delta N=60$).

The corresponding scalar potential has the Starobinsky-like structure,
\begin{equation}
    V_{\alpha}=A\left(1-e^{\sqrt{\frac{2}{\alpha}}\varphi}\right)^2~,\label{V_attractor}
\end{equation}
assuming the canonical kinetic term for $\varphi$. The parameter $A$ is some constant proportional to the inflaton mass squared. In our notation $\alpha$ has a slightly different normalization compared to the original paper \cite{Kallosh:2013yoa} (related by a factor of three).

$\alpha=3$ corresponds to the Starobinsky $f(R)$-gravity model \cite{Starobinsky:1980te}, which can be embedded in higher-derivative supergravity \cite{Ketov:2012yz,Ferrara:2013rsa,Farakos:2013cqa,Ketov:2013dfa,Antoniadis:2014oya,Dalianis:2014aya,Hasegawa:2015era} as well. The correspondence with modified (super)gravity is not shared with the $\alpha\neq 3$ cases.

\begin{figure}
\centering
  \includegraphics[width=.75\linewidth]{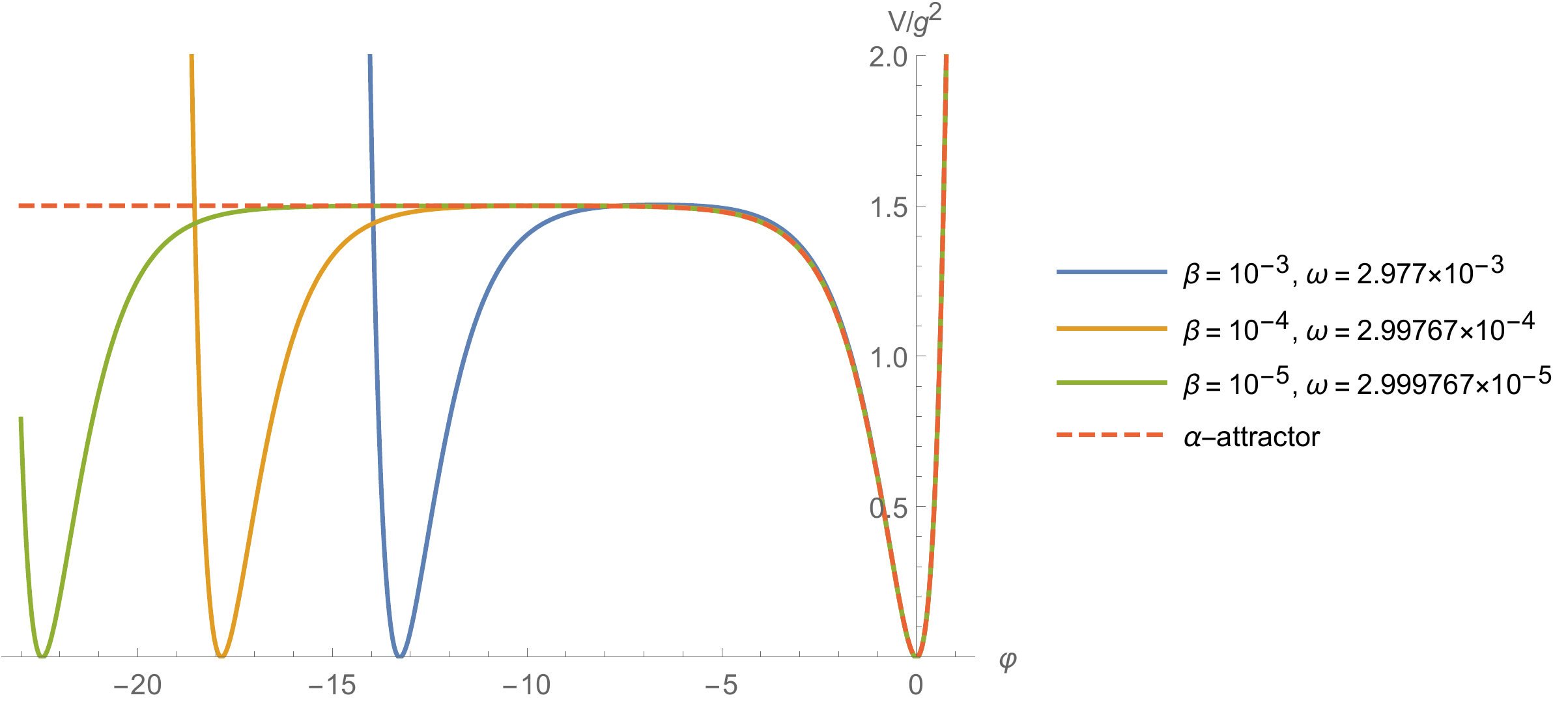}
\captionsetup{width=.9\linewidth}
\caption{The scalar potential \eqref{VF_beta}$+$\eqref{VD_beta} for the canonical inflaton $\varphi$ ($\alpha=2$) and different values of $\beta$ and $\omega$, compared to the $\alpha$-attractor \eqref{V_alpha2_attractor}. The canonical inflaton is shifted, $\varphi\rightarrow\varphi-|\varphi_0|$, so that the minimum is at $\varphi=0$.}
\label{Fig_alpha2_V_attr}
\end{figure}

Going back to our models, let us consider the behavior of the scalar potential \eqref{VF_beta}$+$\eqref{VD_beta} in terms of the canonical scalar, as we change $\beta$. Consider for example $\beta=10^{-3},10^{-4},10^{-5}$ and the corresponding values of $\omega$ found from the vacuum solution of Figure \ref{Fig_alpha2_omega_beta}. The potentials are shown in Figure \ref{Fig_alpha2_V_attr}, where the canonical scalar is shifted as $\varphi\rightarrow\varphi-|\varphi_0|$ where $\varphi_0$ is its VEV, so that $\varphi=0$ coincides with one of the two (near-)Minkowski minima. For reference we included the $\alpha$-attractor potential \eqref{V_attractor} with $\alpha=2$ and $A=3g^2/2$, i.e.
\begin{equation}
    V_{\alpha=2}=\tfrac{3}{2}g^2(1-e^\varphi)^2~.\label{V_alpha2_attractor}
\end{equation}

Remarkably, as $\beta$ and $\omega$ become smaller, we can see that the shape of the potential asymptotically approaches the simple form \eqref{V_alpha2_attractor}. Quantitatively this observation is supported by the comparison of $n_s$, $r$, and the inflaton mass $m_\varphi$ between our potential and the potential \eqref{V_alpha2_attractor} -- see below.

\subsection{SUSY and \texorpdfstring{$R$}{}-symmetry breaking}

Taking four different values of $\beta$ as mentioned above, we numerically estimate the inflationary observables $n_s$, $r$ (using equations of motion), the masses of the inflaton $\varphi$, the vector $A_m$, and the gravitino {\it around the Minkowski vacuum}, as well as the supersymmetry breaking parameters $\langle F\rangle$ and $\langle D\rangle$ given by Eqs.\,\eqref{F_common} and \eqref{D_common} in Appendix. The results are summarized in Table \ref{Tab_alpha2} where we also included the estimates for $n_s$, $r$, and $m_\varphi$ for the reference $\alpha$-attractor potential \eqref{V_alpha2_attractor} (as we are not considering any particular embedding of the model \eqref{V_alpha2_attractor} in supergravity, SUSY breaking parameters are ignored in this case). The first value $\beta=0.005413$ is the marginal case where the lower bound of $n_s^{\rm max}\geq 0.9649$ is saturated (see Figure \ref{Fig_alpha2_omega_beta}), however, as can be seen from Table \ref{Tab_alpha2}, the actual value of $n_s$ at the horizon exit is noticeably smaller, although is still within $2\sigma$ confidence level. Table \ref{Tab_alpha2} also shows that for $\beta\lesssim 10^{-4}$, the predictions of our model for $n_s$ and $r$, as well as the parameters $g$ and $m_\varphi$ practically coincide with the $\alpha$-attractor potential \eqref{V_alpha2_attractor}. 

As for the inflationary Hubble scale, in all the examples it is of the order $10^{13}$ GeV, whereas the {\it effective} masses of the vector and the gravitino, evaluated at the horizon exit, vary from $10^{12}$ GeV when $\beta=0.005413$ to $10^{15}$ GeV when $\beta=10^{-5}$ (both effective masses are of the same order).

\begin{table}[ht]
\centering
\begin{tabular}{l r r r r r}
\toprule
 & $\beta = 0.005413$ & $\beta = 10^{-3}$ & $\beta = 10^{-4}$ & $\beta = 10^{-5}$ & $V_{\alpha=2}$ \\
\hline
 $n_s$ & $0.9528$ & $0.9638$ & $0.9670$ & $0.9671$ & $0.9671$ \\
$r$ & $0.0009$ & $0.0018$ & $0.0021$ & $0.0021$ & $0.0021$ \\
$g$ & $4.11\times 10^{-6}$ & $5.96\times 10^{-6}$ & $6.44\times 10^{-6}$ & $6.45\times 10^{-6}$ & $6.45\times 10^{-6}$ \\
$m_\varphi$ & $1.78\times 10^{13}$ & $2.53\times 10^{13}$ & $2.72\times 10^{13}$ & $2.72\times 10^{13}$ & $2.72\times 10^{13}$ \\
$m_A$ & $7.46\times 10^{14}$ & $5.53\times 10^{15}$ & $5.89\times 10^{16}$ & $5.89\times 10^{17}$ & -- \\
$m_{3/2}$ & $5.28\times 10^{14}$ & $3.91\times 10^{15}$ & $4.16\times 10^{16}$ & $4.17\times 10^{17}$ & -- \\
$|\langle F\rangle|^{1/2}$ & $5.83\times 10^{15}$ & $7.06\times 10^{15}$ & $7.35\times 10^{15}$ & $7.36\times 10^{15}$ & -- \\
$|\langle D\rangle|^{1/2}$ & $4.26\times 10^{16}$ & $1.16\times 10^{17}$ & $3.79\times 10^{17}$ & $1.20\times 10^{18}$ & -- \\\bottomrule
\hline
\end{tabular}
\captionsetup{width=.9\linewidth}
\caption{The inflationary observables $n_s$ and $r$; the inflaton, vector, gravitino masses ($m_\varphi$, $m_A$, $m_{3/2}$, respectively), and SUSY breaking parameters -- all evaluated for four different choices of $\beta$. The observables and the inflaton mass for the $\alpha$-attractor potential \eqref{V_alpha2_attractor} are also included. The parameters $m_\varphi$, $m_A$, $m_{3/2}$, $|\langle F\rangle|^{1/2}$, and $|\langle D\rangle|^{1/2}$ are in units of GeV, while $g$ is dimensionless.}
\label{Tab_alpha2}
\end{table} 

Let us comment on the results of Table \ref{Tab_alpha2}. For smaller values of $\beta$ the inflaton mass can be seen approaching the $\alpha$-attractor value, which can be obtained from Eq.\,\eqref{V_alpha2_attractor} as $m_\varphi=\sqrt{3}g$, yielding $2.72\times 10^{13}~{\rm GeV}$ after using the explicit value of $g$ and restoring the Planck mass.

The mass of the $U(1)_R$ vector field reads
\begin{equation}
    m_A=\sqrt{2}g\langle K_{Z\overbar{Z}}Z\overbar{Z}\rangle^{1/2}~.\label{vector_mass}
\end{equation}
In Figure \ref{Fig_alpha2_omega_beta} we can see that for the interesting values of $\beta$ we have $z_0\equiv\langle Z\rangle\rightarrow 1$ (as $\beta$ decreases), so that $m_A$ is controlled by the VEV of the K\"ahler metric,
\begin{equation}
    K_{Z\overbar{Z}}=\alpha\frac{1+4\beta |Z|^2-\beta |Z|^4}{(1-|Z|^2-\beta |Z|^4)^2}~,
\end{equation}
when varying $\beta$. As $\beta$ decreases and $z_0$ approaches unity, the K\"ahler metric blows up, and so does the vector mass.

The same is true for the gravitino mass
\begin{equation}
    m_{3/2}=\mu\langle e^K Z\overbar{Z}\rangle^{1/2}~,
\end{equation}
and the $D$-field,
\begin{equation}
    \langle D\rangle=-g\frac{1+z_0^2+3\beta z_0^4}{1-z_0^2-\beta z_0^4}
\end{equation}
because $\langle e^K\rangle=(1-z_0^2-\beta z_0^4)^{-2}$ and $(1-z_0^2-\beta z_0^4)^{-1}$ also grow at smaller values of $\beta$.

However, the situation is different for the $F$-field,
\begin{equation}
    \langle F\rangle=-\langle e^{K/2}K^{Z\overbar{Z}}D_{\overbar{Z}}\overbar{W}\rangle=-\frac{\mu}{2}\cdot\frac{1+z_0^2+3\beta z_0^4}{1+4\beta z_0^2-\beta z_0^4}~,
\end{equation}
where the growing factors cancel out and $|\langle F\rangle|^{1/2}$ approaches the constant value $\sim 7.4\times 10^{15}$ GeV.

It can be extrapolated from Table \ref{Tab_alpha2} (and we indeed confirm) that at $\beta=10^{-6}$ the masses $m_A$, $m_{3/2}$, as well as the parameter $|\langle D\rangle|^{1/2}$ reach the Planck scale. Since K\"ahler potential is not protected from quantum corrections, it makes sense to avoid approaching the Planck scale (e.g. $m_{3/2}\ll M_P$), which constrains $\beta$ from below.

\section{\texorpdfstring{$\alpha=3$}{} case}\label{sec_alpha_3}

As in the $\alpha=2$ case, when $\alpha=3$ there is a possibility of a flat de Sitter potential \cite{Aldabergenov:2019ngl}. This time the requirement is $\mu^2=9g^2/2$ which leads to the potential $V=2g^2$ (using the action \eqref{L_beta}--\eqref{VD_beta} with $\beta=0$). Thus we can use the same method of introducing a small deviation from perfect de Sitter model, in order to obtain realistic inflationary scenario. The deviation from the de Sitter relation is measured by the new parameter $\lambda$ as follows,
\begin{equation}
    \mu^2\equiv\tfrac{9}{2}g^2(1+\lambda)~.
\end{equation}
As before, a small non-zero $\beta$ is needed to obtain a stable Minkowski or de Sitter minimum after inflation.

\subsection{The relevant parameter region}

\begin{figure}
\centering
\begin{subfigure}{.49\textwidth}
  \centering
  \includegraphics[width=.9\linewidth]{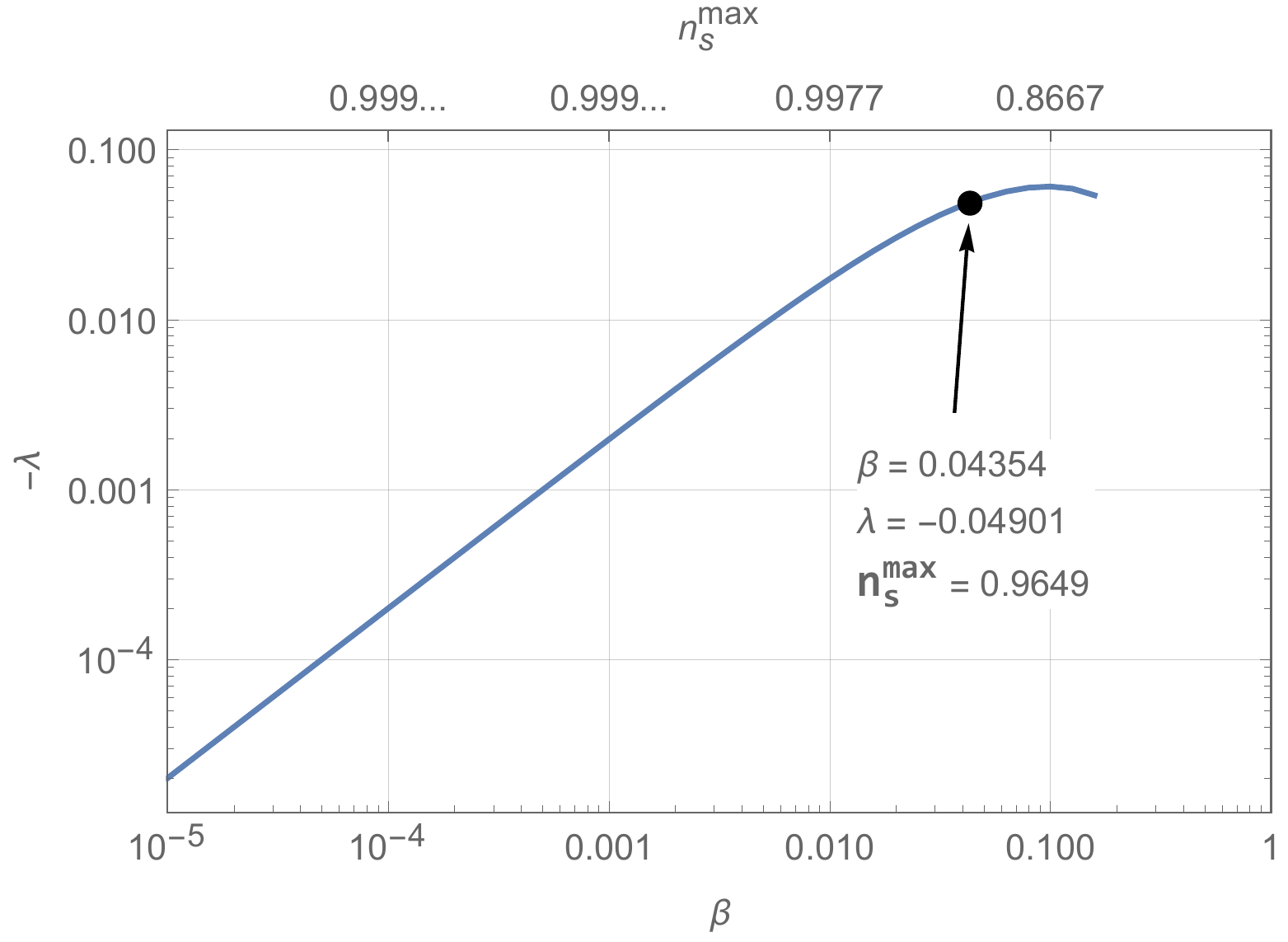}
\end{subfigure}
\begin{subfigure}{.49\textwidth}
  \centering
  \includegraphics[width=.88\linewidth]{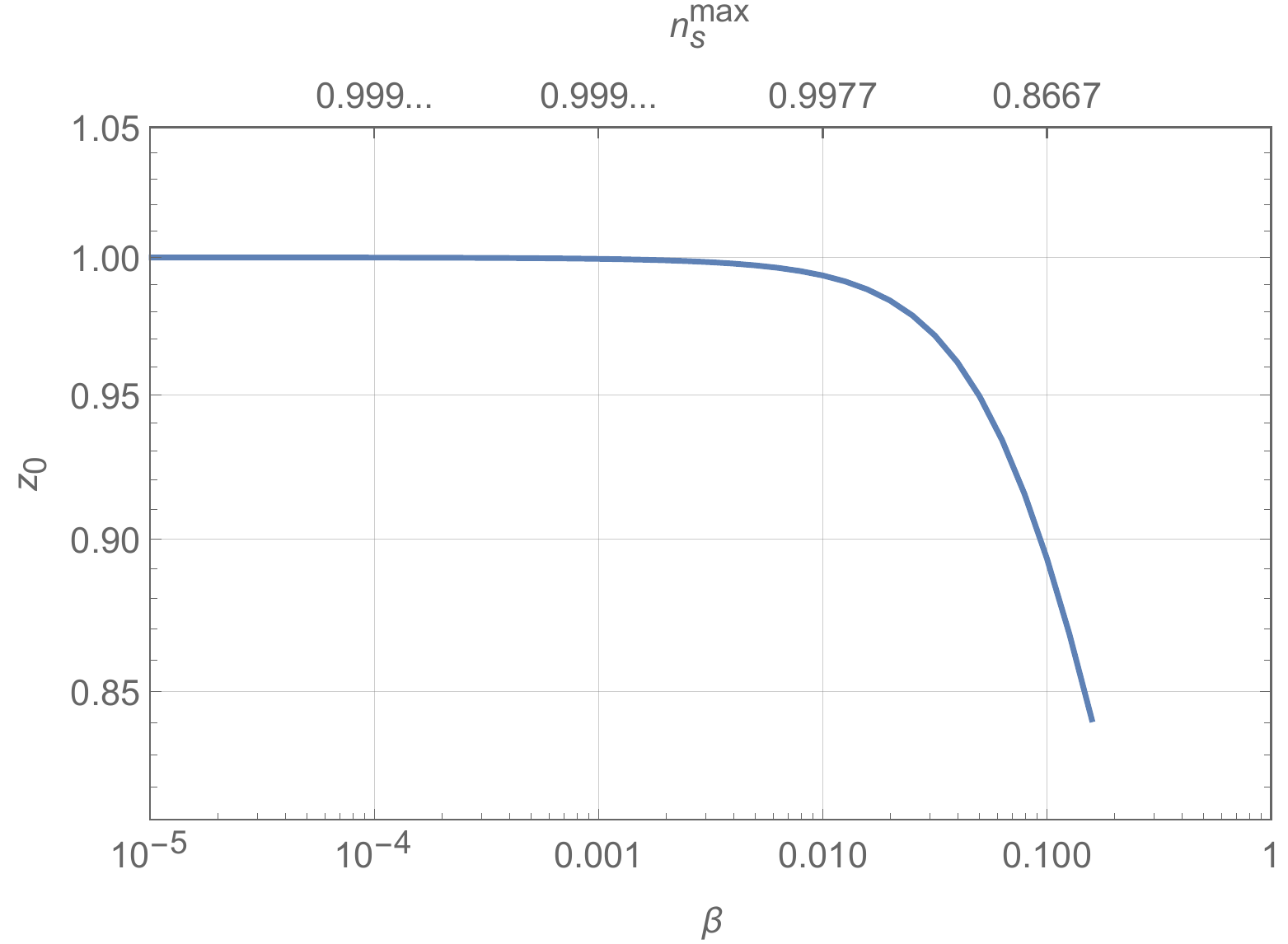}
\end{subfigure}
\captionsetup{width=.9\linewidth}
\caption{Solution to the vacuum equations $V_z=V=0$ ($\alpha=3$ case) with positive $\beta$: $-\lambda(\beta)$ (left) and $z_0(\beta)$ (right). At the top-side of the plots we provide the values of $n_s^{\rm max}$ at reference values of $\beta$.}
\label{Fig_alpha3_lambda_beta_1}
\end{figure}

\begin{figure}
\centering
\begin{subfigure}{.49\textwidth}
  \centering
  \includegraphics[width=.9\linewidth]{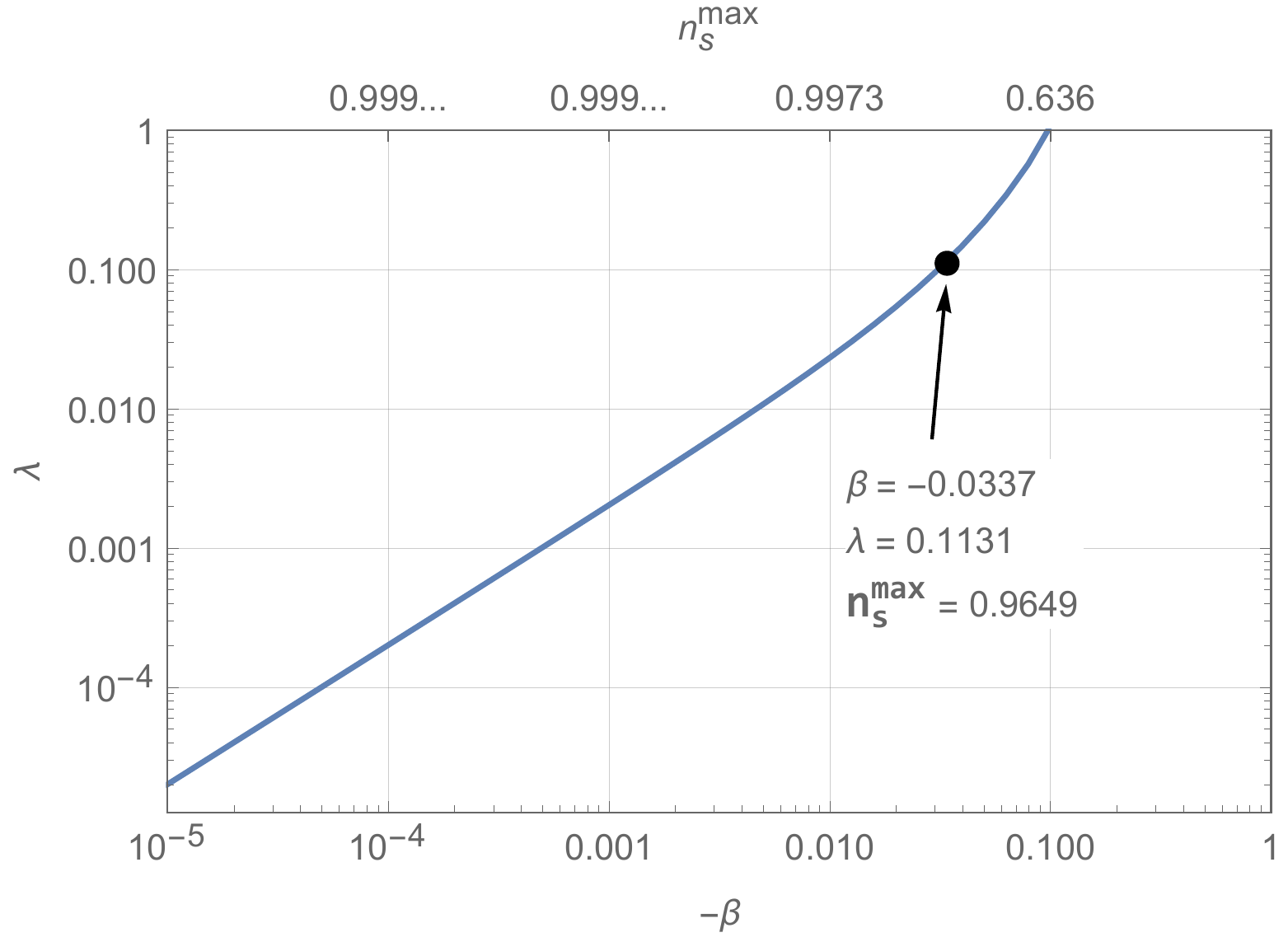}
\end{subfigure}
\begin{subfigure}{.49\textwidth}
  \centering
  \includegraphics[width=.88\linewidth]{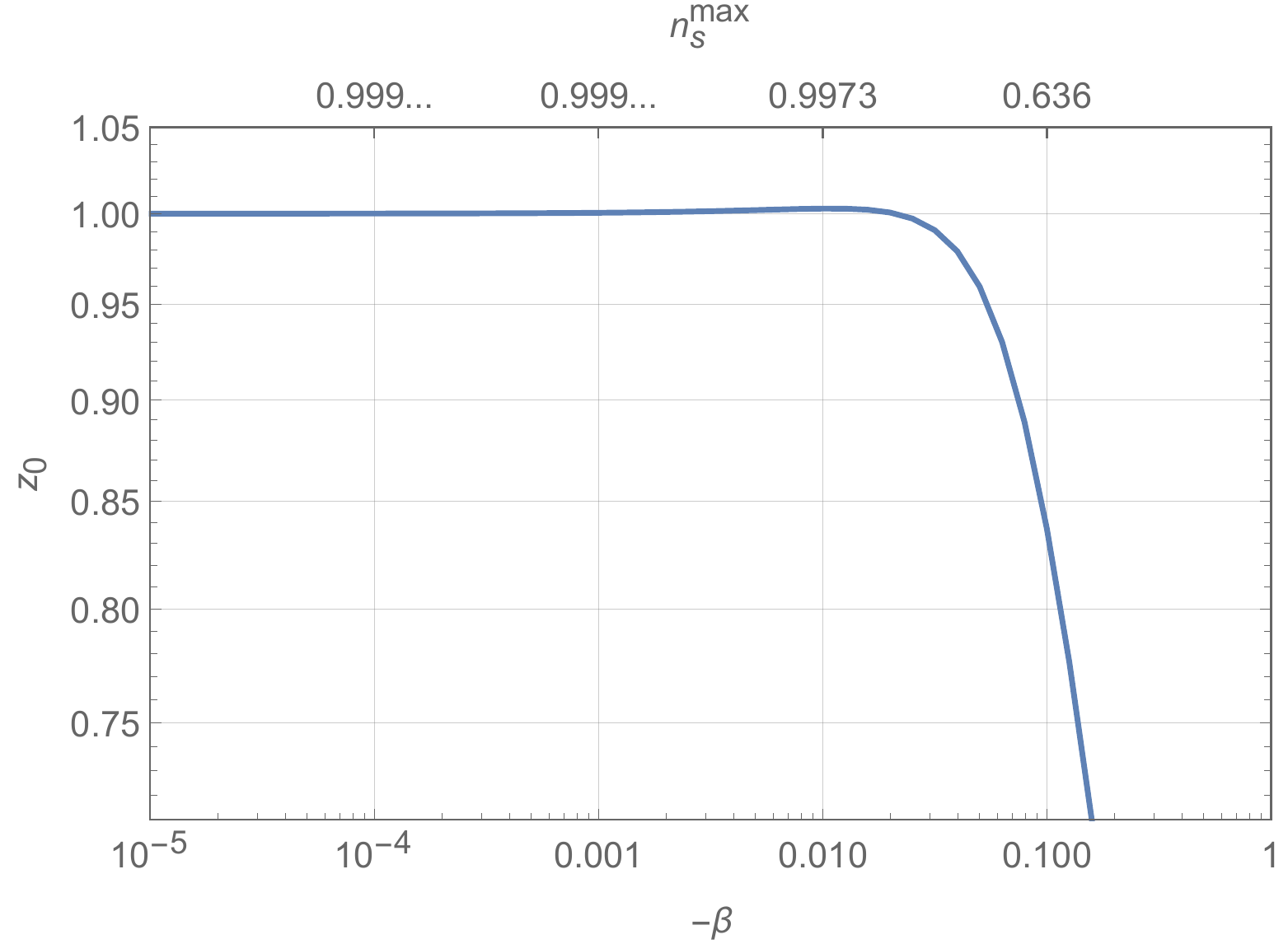}
\end{subfigure}
\captionsetup{width=.9\linewidth}
\caption{Solution to the vacuum equations $V_z=V=0$ ($\alpha=3$ case) with negative $\beta$: $\lambda(-\beta)$ (left) and $z_0(-\beta)$ (right).}
\label{Fig_alpha3_lambda_beta_2}
\end{figure}

To identify the suitable parameter region we again solve the vacuum equations $V_z=V=0$ by varying $\beta$. In contrast to the $\alpha=2$ case, here we find two sets of solutions -- one for positive $\beta$ (and negative $\lambda$) and one for negative $\beta$ (and positive $\lambda$). The solution for positive $\beta$ is shown in Figure \ref{Fig_alpha3_lambda_beta_1}, and the solution for negative $\beta$ in Figure \ref{Fig_alpha3_lambda_beta_2}.

The positive $\beta$ solution of Figure \ref{Fig_alpha3_lambda_beta_1} is cut off at around $\beta\sim 0.2$, indicating that the solution with the assumptions $\beta>0,\lambda<0$ (and $z_0\in\mathbb{R}$) does not exist for larger $\beta$. However this does not affect the relevant (for inflation) part of the solution because it has the upper limit already at $\beta=0.04354$. The negative $\beta$ solution (Figure \ref{Fig_alpha3_lambda_beta_2}) is bounded as $|\beta|\leq 0.0337$.

\subsection{Scalar potential and relation to \texorpdfstring{$\alpha$}{}-attractors}

\begin{figure}
\centering
  \includegraphics[width=.75\linewidth]{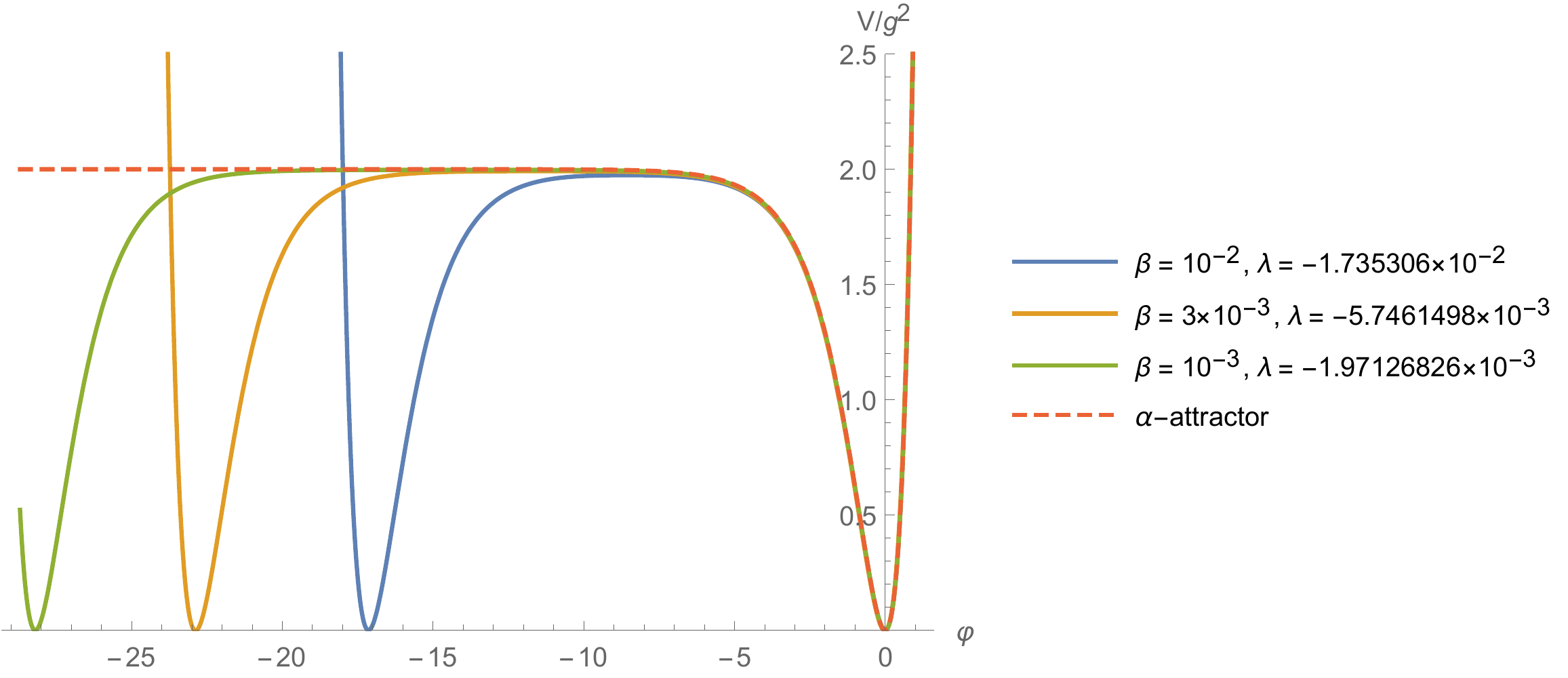}
\captionsetup{width=.9\linewidth}
\caption{The scalar potential \eqref{VF_beta}\eqref{VD_beta} for the canonical inflaton $\varphi$ ($\alpha=3$, positive $\beta$) and different choices of $\beta$ and $\lambda$, compared to the $\alpha$-attractor potential \eqref{V_alpha3_attractor}.}
\label{Fig_alpha3_V_attr_1}
\centering
  \includegraphics[width=.75\linewidth]{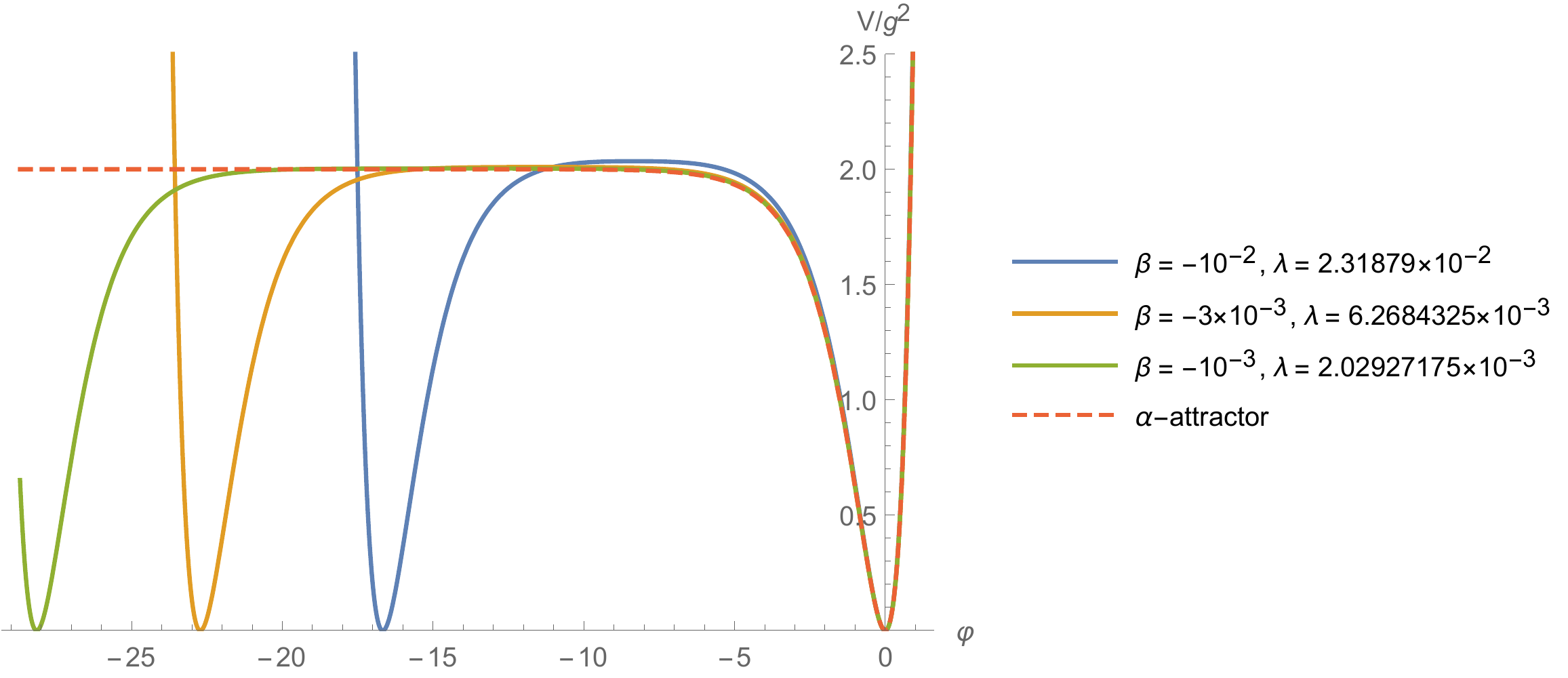}
\captionsetup{width=.9\linewidth}
\caption{The scalar potential \eqref{VF_beta}\eqref{VD_beta} for the canonical inflaton $\varphi$ ($\alpha=3$, negative $\beta$) and different choices of $\beta$ and $\lambda$, compared to the $\alpha$-attractor potential \eqref{V_alpha3_attractor}.}
\label{Fig_alpha3_V_attr_2}
\end{figure}

For $\alpha=3$ we again find that the scalar potential (of the canonically normalized inflaton) asymptotically approaches the corresponding $\alpha$-attractor shape, which for $\alpha=3$ is given by the Starobinsky potential
\begin{equation}
    V_{\alpha=3}=2g^2\left(1-e^{\sqrt{\frac{2}{3}}\varphi}\right)^2~.\label{V_alpha3_attractor}
\end{equation}
To demonstrate this we plot the potential for our model after numerical canonical normalization of the inflaton, with $\alpha=3$ and different choices of $\beta$ and $\lambda$, and compare it with the potential \eqref{V_alpha3_attractor}. The plot for positive $\beta$ is shown in Figure \ref{Fig_alpha3_V_attr_1}, and the plot for negative $\beta$ in Figure \ref{Fig_alpha3_V_attr_2}.

\subsection{SUSY and R-symmetry breaking}

Here we summarize the results for the inflationary observables $n_s$ and $r$ (found from the equations of motion), the masses $m_{\varphi}$, $m_A$, and $m_{3/2}$, and the SUSY breaking parameters $\langle F\rangle$, $\langle D\rangle$, for various choices of $\beta$ (the masses and SUSY breaking parameters are evaluated around the Minkowski vacuum). The values of the gauge coupling $g$ in each case are set by the observed value of the scalar amplitude \eqref{nsrA_obs}.

\begin{table}[hbt!]
\centering
\begin{tabular}{l r r r r}
\toprule
 & $\beta = 0.04354$ & $\beta = 10^{-2}$ & $\beta = 3\times 10^{-3}$ & $V_{\alpha=3}$ \\
\hline
$n_s$ & $0.9517$ & $0.9659$ & $0.9673$ & $0.9674$ \\
$r$ & $0.0010$ & $0.0029$ & $0.0031$ & $0.0031$ \\
$g$ & $3.85\times 10^{-6}$ & $6.60\times 10^{-6}$ & $6.77\times 10^{-6}$ & $6.76\times 10^{-6}$ \\
$m_\varphi$ & $1.57\times 10^{13}$ & $2.62\times 10^{13}$ & $2.69\times 10^{13}$ & $2.69\times 10^{13}$ \\
$m_A$ & $4.89\times 10^{14}$ & $1.08\times 10^{16}$ & $1.15\times 10^{17}$ & -- \\
$m_{3/2}$ & $1.79\times 10^{15}$ & $1.52\times 10^{17}$ & $5.25\times 10^{18}$ & -- \\
$|\langle F\rangle|^{1/2}$ & $1.39\times 10^{16}$ & $3.65\times 10^{16}$ & $6.70\times 10^{16}$ & -- \\
$|\langle D\rangle|^{1/2}$ & $3.80\times 10^{16}$ & $1.80\times 10^{17}$ & $5.85\times 10^{17}$ & -- \\\bottomrule
\hline
\end{tabular}
\captionsetup{width=.9\linewidth}
\caption{The inflationary observables $n_s$ and $r$; the inflaton, vector, gravitino masses ($m_\varphi$, $m_A$, $m_{3/2}$, respectively), and SUSY breaking parameters for our model with $\alpha=3$ and positive $\beta$. The observables and the inflaton mass for the $\alpha$-attractor potential \eqref{V_alpha3_attractor} are included. The parameters $m_\varphi$, $m_A$, $m_{3/2}$, $|\langle F\rangle|^{1/2}$, and $|\langle D\rangle|^{1/2}$ are in units of GeV, while $g$ is dimensionless.}
\label{Tab_alpha3_1}

\bigskip

\begin{tabular}{l r r r r}
\toprule
 & $\beta = -0.0337$ & $\beta = -10^{-2}$ & $\beta = -3\times 10^{-3}$ & $V_{\alpha=3}$ \\
\hline
$n_s$ & $0.9521$ & $0.9652$ & $0.9673$ & $0.9674$ \\
$r$ & $0.0008$ & $0.0028$ & $0.0031$ & $0.0031$ \\
$g$ & $3.29\times 10^{-6}$ & $6.41\times 10^{-6}$ & $6.74\times 10^{-6}$ & $6.76\times 10^{-6}$ \\
$m_\varphi$ & $1.45\times 10^{13}$ & $2.58\times 10^{13}$ & $2.69\times 10^{13}$ & $2.69\times 10^{13}$ \\
$m_A$ & $3.29\times 10^{14}$ & $8.55\times 10^{15}$ & $1.07\times 10^{17}$ & -- \\
$m_{3/2}$ & $1.34\times 10^{15}$ & $1.15\times 10^{17}$ & $4.82\times 10^{18}$ & -- \\
$|\langle F\rangle|^{1/2}$ & $1.38\times 10^{16}$ & $3.53\times 10^{16}$ & $6.65\times 10^{16}$ & -- \\
$|\langle D\rangle|^{1/2}$ & $3.12\times 10^{16}$ & $1.60\times 10^{17}$ & $5.65\times 10^{17}$ & -- \\\bottomrule
\hline
\end{tabular}
\captionsetup{width=.9\linewidth}
\caption{The inflationary observables, relevant masses, and SUSY breaking parameters for $\alpha=3$ and negative $\beta$. The parameters $m_\varphi$, $m_A$, $m_{3/2}$, $|\langle F\rangle|^{1/2}$, and $|\langle D\rangle|^{1/2}$ are in units of GeV, while $g$ is dimensionless.}
\label{Tab_alpha3_2}
\end{table}

In Tables \ref{Tab_alpha3_1} and \ref{Tab_alpha3_2} the cases of positive and negative $\beta$ are considered.\footnote{$\beta=0.04354$ and $\beta=-0.0337$ are the marginal cases taken from the vacuum solutions of Figures \ref{Fig_alpha3_lambda_beta_1} and \ref{Fig_alpha3_lambda_beta_2}, respectively.} Like in the $\alpha=2$ case, Figures \ref{Fig_alpha3_lambda_beta_1} and \ref{Fig_alpha3_lambda_beta_2} show that the smaller the value of $|\beta|$ is, the closer $z_0$ is to unity. Therefore $m_{3/2}$ and $m_A$ become larger proportionally to
\begin{equation}
    \langle e^{K/2}\rangle=(1-z_0^2-\beta z_0^4)^{-3/2}~,~~~\langle K_{Z\overbar{Z}}\rangle^{1/2}=\frac{\sqrt{3(1+4\beta z_0^2-\beta z_0^4)}}{1-z_0^2-\beta z_0^4}~,
\end{equation}
respectively. The vacuum values of $F$- and $D$-fields are given by
\begin{equation}
    \langle F\rangle=-\frac{\mu}{3}\cdot\frac{1+2z_0^2+5\beta z_0^4}{(1+4\beta z_0^2-\beta z_0^4)\sqrt{1-z_0^2-\beta z_0^4}}~,~~~\langle D\rangle=-g\frac{1+2z_0^2+5\beta z_0^4}{1-z_0^2-\beta z_0^4}~,
\end{equation}
where in contrast to the $\alpha=2$ case, the growing factor in $\langle F\rangle$ is not fully cancelled when $\alpha=3$, and we have $\langle F\rangle\propto (1-z_0^2-\beta z_0^4)^{-1/2}$ that grows when $\beta$ decreases, albeit at a slower rate than $\langle D\rangle$.

Tables \ref{Tab_alpha3_1} and \ref{Tab_alpha3_2} also show that the predictions of the models with $\alpha=3$ for the inflationary parameters are nearly indistinguishable from the those of the Starobinsky model \eqref{V_alpha3_attractor} already at $|\beta|=3\times 10^{-3}$ (for both positive and negative $\beta$ solutions). However, the gravitino mass in these cases is already reaching the Planck scale, which can be avoided by raising the absolute value of $\beta$ (we will also comment on the lower limit on $|\beta|$ coming from swampland distance conjecture in Section \ref{sec_SDC}).

The Hubble scale is the same as before, at $10^{13}$ GeV. When $\beta=0.04354$ (or $\beta=-0.0337$ for the negative $\beta$ case), the effective masses of the vector and the gravitino at the horizon exit are around $10^{12}$ GeV, while for $|\beta|=10^{-3}$ (for both positive and negative $\beta$ cases) we have $m_A^{\rm eff}\sim 10^{16}$ GeV and $m_{3/2}^{\rm eff}\sim 10^{17}$ GeV.

\section{Fermion spectrum}\label{sec_fermion}

The spectrum of our models is given by three distinct (on-shell) multiplets: supergravitational multiplet $\{e^a_m,\psi_m\}$, chiral multiplet $\{Z,\chi\}$, and (massless) vector multiplet $\{A_m,\lambda\}$, where $\psi_m$ is the gravitino field, $\chi$ and $\lambda$ are spin-$1/2$ fermions (inflatino and gaugino).\footnote{In terms of the $R$-charge $q$ of the superpotential, the $R$-charges of $Z$, $\chi$, $\lambda$, and $\psi_m$ are $q_Z=q$, $q_\chi=q_\lambda=q_\psi=q/2$. We take $q=1$ in our examples.} In the previous sections we described the physical spectrum of bosons which consists of the massive real scalar $|Z|$ and a massive vector that can be thought of as the gauge-invariant combination of $A_m$ and the angular component of $Z$. As we showed, the gravitino acquires non-vanishing mass due to spontaneous breaking of supersymmetry. Thus, the spectrum must include a goldstino that would represent longitudinal modes of the massive gravitino. Since in our models SUSY is broken by both $F$- and $D$-terms, the goldstino $\eta$ must be the following linear combination of $\chi$ and $\lambda$,
\begin{equation}
    \eta={\cal G}_Z\chi-\frac{ig{\mathscr D}}{\sqrt{2}W}e^{-K/2}\lambda~,\label{goldstino}
\end{equation}
where $\cal G$ is the K\"ahler-invariant function ${\cal G}=K+\log|W|^2$, ${\cal G}_Z$ is its derivative with respect to $Z$, and $\mathscr D$ is the Killing potential of $U(1)_R$ gauge symmetry (see Appendix).

When SUSY is spontaneously broken, the (supersymmetric) unitary gauge reveals the physical mass spectrum where $\eta=0$, and the gravitino is massive. Then, the combination of $\chi$ and $\lambda$, orthogonal to the goldstino, becomes a physical massive fermion. After proper normalization of its kinetic term and some manipulations, the mass of this fermion can be written as
\begin{equation}
    m_{1/2}=2m_{3/2}\left\langle\frac{|g^2 Z\overbar{Z}(\partial_Z-\Gamma^Z_{ZZ}+{\cal G}_Z){\cal G}_Z-2m^2_A|}{4m^2_{3/2}+m^2_A}\right\rangle~,
\end{equation}
where $\Gamma^Z_{ZZ}$ is the Christoffel symbol of the K\"ahler manifold, the gravitino mass is $m_{3/2}=\langle e^{{\cal G}/2}\rangle$, and the vector mass $m_A$ is given by Eq.\,\eqref{vector_mass}.

We find that for the $\alpha=2$ model described in Section \ref{sec_alpha_2} we have $m_{1/2}\approx m_{3/2}$ (with $m_{1/2}$ being a bit larger). For the $\alpha=3$ examples described in Tables \ref{Tab_alpha3_1} and \ref{Tab_alpha3_2} the fermion mass $m_{1/2}$ varies from $10^{13}$ to $10^{15}$ GeV (larger for smaller $|\beta|$), being second lightest field of the spectrum, after the inflaton $\varphi$. On the other hand, the effective mass of this fermion at the horizon exit varies from $10^{12}$ to $10^{15}$ GeV (larger for smaller $|\beta|$) for the examples of Tables \ref{Tab_alpha2}, \ref{Tab_alpha3_1}, and \ref{Tab_alpha3_2}.

\section{Comment on Swampland Distance Conjecture}\label{sec_SDC}

Swampland Distance Conjecture (SDC) \cite{Ooguri:2006in} states that scalar fields travelling infinite distances in field space give rise to an infinite tower of massless (Kaluza--Klein, wrapping D-brane, etc.) states, which indicates breakdown of the effective field theory (EFT). When applied to inflationary EFT, the conjecture imposes the following upper limit on the (proper) field-space displacement \cite{Scalisi:2018eaz},
\begin{equation}
    \Delta\varphi_{\rm sdc}=\frac{1}{c}\log\left(\frac{M_P}{H}\right)~,
\end{equation}
where $c$ is some constant of order one, and $H$ is the inflationary Hubble scale.

Let us demonstrate that application of the SDC to our models can further constrain the parameter space. Taking $c=1$ and $H=10^{13}$ GeV leads to $\Delta\varphi_{\rm sdc}=12~M_P$. If we require that our models hold as EFTs all the way up to the symmetric phase where the $U(1)_R$ is restored, SDC puts the upper limit on the distance $\Delta\varphi$ (for the canonical scalar $\varphi$) between the symmetric point (local maximum) and the minimum. In the $\alpha=2$ examples shown in Figure \ref{Fig_alpha2_V_attr}, it can be seen that this imposes the lower bound (roughly) $\beta\gtrsim 10^{-5}$. On the other hand, Figures \ref{Fig_alpha3_V_attr_1} and \ref{Fig_alpha3_V_attr_2} representing the $\alpha=3$ case imply the corresponding lower bound $|\beta|\gtrsim 3\times 10^{-3}$.

Because our models feature a $U(1)_R$ gauge symmetry, one might wonder how they hold up against the Weak Gravity Conjecture \cite{ArkaniHamed:2006dz} that states that a theory with a $U(1)$
gauge symmetry must include at least one particle satisfying
\begin{equation}
    \frac{m}{|qg|}\leq M_P~,\label{WGC}
\end{equation}
where $m$ is the mass of this particle, $q$ is its charge, and $g$ is the gauge coupling. It is not difficult to check that none of the particles of our models satisfy this bound. However when (Supersymmetric) Standard Model superfields are inevitably added, some of their component fields will necessarily carry non-zero $R$-charge and may thus satisfy the bound \eqref{WGC} if their masses are not too large. The details depend on a particular realization of the Standard Model within our framework which is beyond the scope of this work.

\section{Conclusion}\label{sec_concl}

In this work we introduced a new class of inflationary models with spontaneous supersymmetry breaking after inflation, where inflaton is identified with the superpartner of the goldstino. As a starting point we used supersymmetry breaking models introduced in Ref. \cite{Aldabergenov:2019ngl}, utilizing $SU(1,1)/U(1)$ hyperbolic geometry of the K\"ahler manifold and gauged $U(1)_R$ phase symmetry that fixes the superpotential to be a monomial of the chiral scalar $Z$. Choosing the simplest superpotential $W=\mu Z$, we showed that adding only the leading-order (perturbative) correction term to the K\"ahler potential as
\begin{equation}
    K=-\alpha\log(1-|Z|^2-\beta|Z|^4)~,
\end{equation}
is enough to obtain realistic slow-roll inflation when $\alpha=2$ and $\alpha=3$, in contrast to the model of \cite{Antoniadis:2016aal} where {\it non-perturbative} correction is needed (and the K\"ahler potential is of the canonical type). Our models have three parameters (after $\alpha$ is chosen) $\beta$, $\mu$, and the gauge coupling $g$, while it is more convenient to trade a combination of $g$ and $\mu$ for another parameter $\omega$ defined as $1+\omega=\mu^2/(2g^2)$ (or the parameter $\lambda$ defined as $1+\lambda=2\mu^2/(9g^2)$ in the case $\alpha=3$).

We found Minkowski vacuum solutions with a possibility for de Sitter uplifting, and numerically derived the (semi-infinite) trajectories (shown in Figures \mbox{\ref{Fig_alpha2_omega_beta}, \ref{Fig_alpha3_lambda_beta_1} and \ref{Fig_alpha3_lambda_beta_2})} in the $\beta-\omega$ parameter space (or $\beta-\lambda$ space for $\alpha=3$) allowing for such vacua. Both Minkowski and de Sitter vacua require equal amount of fine-tuning of the parameters which means choosing a point exactly along these trajectories. Thus, we can almost freely choose one parameter (as long as it is smaller than a certain value), while the other one must be precisely fixed (up to numerical precision).

We learned that when e.g. $\alpha=2$, slow-roll inflation requires positive non-zero $\beta$ bounded from above as $\beta\lesssim 10^{-3}$, while when $\alpha=3$ both positive and negative $\beta$ are allowed with the bound $|\beta|\lesssim 10^{-2}$. The parameter $\omega$ is also bounded from above as $\omega\lesssim 10^{-2}$ which means the ratio $\mu^2/(2g^2)\approx 1$ (in Planck units), while the magnitude of both $\mu$ and $g$ is fixed by the observations of the amplitude of scalar perturbations. The suitable values of $g$ are given in Tables \ref{Tab_alpha2}, \ref{Tab_alpha3_1}, and \ref{Tab_alpha3_2}. After inflation, supersymmetry and R-symmetry are spontaneously broken by non-vanishing $F$- and $D$-terms, and the physical spectrum of the models (aside from the graviton) is given by the inflaton, a spin-$1/2$ fermion, a vector, and the gravitino, all of which have masses around the inflationary scale or larger (the lightest field being the inflaton).

Remarkably, despite the fact that the canonically parametrized scalar potential in our models is available only numerically (due to the $\beta$-correction term), we found that for small enough but non-zero $|\beta|$, our potentials asymptotically approach simple $\alpha$-attractor E-models, as can be seen from the comparison in Figures \ref{Fig_alpha2_V_attr}, \ref{Fig_alpha3_V_attr_1}, \ref{Fig_alpha3_V_attr_2}, and Tables \ref{Tab_alpha2}, \ref{Tab_alpha3_1}, \ref{Tab_alpha3_2}.

Finally, we showed how Swampland Distance Conjecture can put lower bounds on the parameter $|\beta|$ in both $\alpha=2$ and $\alpha=3$ models, if we require that the models are valid (as EFTs) up to the $U(1)_R$ symmetric phase. In addition, since K\"ahler potential can receive quantum corrections that can spoil fine-tuned parameters, the mass parameters approaching $M_P$ should be avoided in general. This places similar lower limits on $|\beta|$. In summary, $\beta$ and $\omega$ (or $\lambda$) are bounded from above by CMB observations of the spectral tilt, while they are bounded from below by quantum gravity considerations.

As regards the stability of the inflationary solutions and the Minkowski/dS vacua against adding matter fields to the model, we coupled our model to a generic matter field with canonical K\"ahler potential (with no couplings to the inflaton in the K\"ahler potential) and Wess--Zumino superpotential $W_{\rm WZ}=a\Phi^2+b\Phi^3$ that enters the total superpotential as
\begin{equation}
    W=(\mu+W_{\rm WZ})Z~.\label{W_tot}
\end{equation}
This interacting toy model with the vanishing VEV $\langle\Phi\rangle$=0 is compatible with the vacua and inflationary solutions that we obtained in this work. In particular the second derivative of the total potential, $V_{\Phi\overbar{\Phi}}$, is positive at $\langle\Phi\rangle$=0 (both during inflation and after), i.e. no tachyonic instabilities. The reason of choosing the particular form of the total superpotential \eqref{W_tot} is that the $R$-charges of the MSSM superfields can be zero (while their fermionic components will of course carry non-zero $R$-charges). The coupling between $W_{\rm WZ}$ and $Z$ in the superpotential may lead to some interesting phenomenological implications, which can be studied in the future. Of course, if other couplings between the inflaton and matter are present (e.g. in the K\"ahler potential), the stability of the model should be re-evaluated.

In future works it would also be interesting to consider other values of $\alpha$. Since we already know that when $\alpha\neq 2,3$ it proves difficult to achieve viable inflationary model with only the leading-order correction ($\beta |Z|^4$ term), the next step would be to see if adding sixth-order correction would help, and, if the answer is yes, whether or not the $\alpha$-attractor-like behavior would emerge as in the $\alpha=2,3$ models.

\section*{Acknowledgements}

This work was supported by CUniverse research promotion project of Chulalongkorn University in Bangkok, Thailand, under the grant No. CUAASC. Y.A. was supported in part by the Ministry of Education and Science of the Republic of Kazakhstan under the grant No. AP05133630. A.C. was supported by "CU Global Partnership Project" under the grant No. B16F630071.

\section*{Appendix: $N=1$ supergravity in curved superspace}

We use Wess--Bagger conventions \cite{Wess:1992cp}, where the superspace action for the chiral superfield $\bf Z$ coupled to the $U(1)_R$ vector superfield $\bf V$ in standard Poincar\'e supergravity reads ($M_{\rm Pl}=1$)
\begin{equation}
    {\cal L}=\int d^2\Theta 2{\cal E}\left[\tfrac{3}{8}(\overbar{\cal D}^2-8{\cal R})e^{-\tfrac{1}{3}(K+\Gamma)}+W+\tfrac{1}{4}h{\cal W}^\alpha{\cal W}_\alpha\right]+{\rm h.c.}~.\label{App_L_super}
\end{equation}
Here $\cal E$ is the chiral density superfield, $\cal R$ is the chiral curvature superfield, ${\cal D}_\alpha,\overbar{\cal D}_{\dot{\alpha}}$ are the superspace (fermionic) covariant derivatives with ${\cal D}^2\equiv{\cal D}^\alpha{\cal D}_\alpha$ and $\overbar{\cal D}^2\equiv\overbar{\cal D}_{\dot{\alpha}}\overbar{\cal D}^{\dot{\alpha}}$; $K=K({\bf Z},\overbar{\bf Z})$ is the superfield K\"ahler potential and $\Gamma=\Gamma({\bf Z},\overbar{\bf Z},{\bf V})$ is the compensating term for the gauge transformation of $K$; $W=W({\bf Z})$ is the superpotential, $h=h({\bf Z})$ is the gauge kinetic function (which is not gauge invariant on its own, but can be used for the cancellation of possible one-loop anomalies), and ${\cal W}_\alpha\equiv -\tfrac{1}{4}(\overbar{\cal D}^2-8{\cal R}){\cal D}_\alpha{\bf V}$ is the superfield strength of $\bf V$. The operator $(\overbar{\cal D}^2-8{\cal R})$ is the chiral projector in curved superspace. The function $\Gamma$ can be expanded in terms of $\bf V$ as
\begin{equation}
    \Gamma=2g\mathscr{D}{\bf V}+2g^2K_{Z\overbar{Z}}X^Z\overbar{X}^{\overbar{Z}}{\bf V}^2+{\cal O}({\bf V}^3)~,
\end{equation}
where the higher order terms (starting from cubic) vanish in Wess--Zumino gauge. Here $\mathscr{D}$ is the Killing potential of the $U(1)_R$ gauge symmetry, $X^Z$ is the Killing vector, $g$ is the gauge coupling, and $K_{Z\overbar{Z}}\equiv\partial_{\bf Z}\partial_{\overbar{\bf Z}}K$ is the K\"ahler metric. Killing potential can be written as
\begin{equation}
\mathscr{D}=iX^Z\left(K_Z+\frac{W_Z}{W}\right)~,
\end{equation}
where the superpotential-dependent term is the Fayet-Iliopoulos term of gauged $R$-symmetry.

The superfields $\bf Z$ and $\bf V$ can be expanded as (using Wess--Zumino gauge)
\begin{gather}
    {\bf Z}=Z+\sqrt{2}\Theta\chi+\Theta^2{\cal F}~,\label{Z_expansion}\\
    {\bf V}=-\Theta\sigma^m\overbar{\Theta}A_m+i\Theta^2\overbar{\Theta}\overbar{\lambda}-i\overbar{\Theta}^2\Theta\lambda+\tfrac{1}{2}\Theta^2\overbar{\Theta}^2d~,\label{V_expansion}
\end{gather}
where $Z$ is complex scalar, $\chi$ and $\lambda$ are spin-$1/2$ (Weyl) fermions, $A_m$ is the $U(1)_R$ gauge field, $\cal F$ (complex) and $d$ (real) are auxiliary scalars.

After expanding the superspace Lagrangian \eqref{App_L_super} in the component fields, the auxiliary fields can be eliminated via their equations of motion. In particular, for ${\cal F}$ and $d$ we have
\begin{gather}
    {\cal F}=-e^{K/3}K^{Z\overbar{Z}}D_{\overbar{Z}}\overbar{W}~,\label{F_EOM}\\
    d=-ge^{-K/3}\mathscr{D}~,\label{D_EOM}
\end{gather}
where the leading components ($\Theta=0$) are assumed on the right-hand side. Then, after proper Weyl rescaling we arrive at the Lagrangian (keeping only the bosonic terms),
\begin{equation}
    e^{-1}{\cal L}=\tfrac{1}{2}R-K_{Z\overbar{Z}}D_mZ\overbar{D^mZ}-\tfrac{1}{4}{\rm Re}(h)F_{mn}F^{mn}+\tfrac{1}{8}{\rm Im}(h)\epsilon^{mnkl}F_{mn}F_{kl}-V_F-V_D~,\label{App_L_comp}
\end{equation}
with the scalar potential given by
\begin{gather}
    V_F=e^K\left(K^{Z\overbar{Z}}D_ZW D_{\overbar{Z}}\overbar{W}-3|W|^2\right)~,\label{App_VF}\\
    V_D=\frac{g^2}{2}{\rm Re}(h)^{-1}\mathscr{D}^2~,
\end{gather}
where the following notation is used,
\begin{gather}
    K^{Z\overbar{Z}}\equiv K^{-1}_{Z\overbar{Z}}~,~~~D_ZW\equiv\frac{\partial W}{\partial Z}+W\frac{\partial K}{\partial Z}~.
\end{gather}
The covariant derivative of $Z$ is defined as
\begin{equation}
    D_mZ\equiv\partial_m Z-gA_m X^Z~.
\end{equation}
Since we consider the $U(1)_R$ as the phase symmetry $Z\rightarrow e^{-iq_Z\theta}Z$ ($q_Z$ -- R-charge of $Z$, $\theta$ -- gauge parameter), the Killing vector takes the form $X^Z=-iq_ZZ$.

It should be noted that there is a difference between the auxiliary ${\cal F}$- and $d$-fields defined as the expansion coefficients in Eqs.\,\eqref{Z_expansion} and \eqref{V_expansion}, and the more commonly used (rescaled) definitions
\begin{gather}
    F=-e^{K/2}K^{Z\overbar{Z}}D_{\overbar{Z}}\overbar{W}~,\label{F_common}\\
    D=-g\mathscr{D}~,\label{D_common}
\end{gather}
which are convenient because the scalar potential can be written as
\begin{equation}
    V_F=K_{Z\overbar{Z}}|F|^2-3m_{3/2}^2~,~~~V_D=\tfrac{1}{2}{\rm Re}(h)^{-1}D^2~,\label{V_FD_app}
\end{equation}
similarly to the case of global supersymmetry (except for the gravitino mass term),
whereas if we use ${\cal F}$ and $d$, extra $K$-dependent factors will appear,
\begin{equation}
    V_F=e^{K/3}K_{Z\overbar{Z}}|{\cal F}|^2-3m^2_{3/2}~,~~~V_D=\tfrac{1}{2}{\rm Re}(h)^{-1}e^{2K/3}\,d^2~.
\end{equation}
This difference between $\cal F$, $d$ and $F$, $D$ is related to the fact that the component form of the Lagrangian \eqref{App_L_super} is originally in the Jordan frame, and the extra $K$-dependent factors are brought in by Weyl rescaling when going to the Einstein frame.

The two definitions of the auxiliary fields are related by 
\begin{equation}
    {\cal F}=e^{-K/6}F~,~~~d=e^{-K/3}D~.
\end{equation}
In this work we use the common definitions $F$, $D$ and their vacuum expectation values, for the parametrization of supersymmetry breaking scale.

\providecommand{\href}[2]{#2}\begingroup\raggedright\endgroup

\end{document}